\pdfoutput=1
\documentclass[12pt]{article}

\usepackage[pdftex]{graphicx}
\usepackage{graphicx}
\usepackage{subcaption}
\usepackage{colordvi}
\usepackage{fancybox}
\usepackage{cancel}
\usepackage{amsmath}
\usepackage{amsfonts}
\usepackage{slashed}
\usepackage{amssymb}
\usepackage{caption}
\usepackage{appendix}
\usepackage{hyperref} 
\usepackage{multirow}
\hypersetup{colorlinks=true, linkcolor=black, urlcolor=blue}
\usepackage{float}
\usepackage{cite}
\usepackage{mathtools}
\usepackage{bbm}

\begin{document}

\newcommand{\ba}[1]{\begin{array}{#1}} \newcommand{\ea}{\end{array}}

\numberwithin{equation}{section}


\def\Journal#1#2#3#4{{#1} {\bf #2}, #3 (#4)}

\def\NCA{\em Nuovo Cimento}
\def\NIM{\em Nucl. Instrum. Methods}
\def\NIMA{{\em Nucl. Instrum. Methods} A}
\def\NPB{{\em Nucl. Phys.} B}
\def\PLB{{\em Phys. Lett.}  B}
\def\PRL{\em Phys. Rev. Lett.}
\def\PRD{{\em Phys. Rev.} D}
\def\ZPC{{\em Z. Phys.} C}

\def\st{\scriptstyle}
\def\sst{\scriptscriptstyle}
\def\mco{\multicolumn}
\def\epp{\epsilon^{\prime}}
\def\vep{\varepsilon}
\def\ra{\rightarrow}
\def\ppg{\pi^+\pi^-\gamma}
\def\vp{{\bf p}}
\def\ko{K^0}
\def\kb{\bar{K^0}}
\def\al{\alpha}
\def\ab{\bar{\alpha}}

\def\np{Nucl. Phys. {\bf B}}\def\pl{Phys. Lett. {\bf B}}
\def\mpl{Mod. Phys. {\bf A}}\def\ijmp{Int. J. Mod. Phys. {\bf A}}
\def\cmp{Comm. Math. Phys.}\def\prd{Phys. Rev. {\bf D}}

\def\oa{\bigcirc\!\!\!\! a}
\def\ob{\bigcirc\!\!\!\! b}
\def\oc{\bigcirc\!\!\!\! c}
\def\oi{\bigcirc\!\!\!\! i}
\def\oj{\bigcirc\!\!\!\! j}
\def\ok{\bigcirc\!\!\!\! k}
\def\ve{\vec e}\def\vk{\vec k}\def\vn{\vec n}\def\vp{\vec p}
\def\vr{\vec r}\def\vs{\vec s}\def\vt{\vec t}\def\vu{\vec u}
\def\vv{\vec v}\def\vx{\vec x}\def\vy{\vec y}\def\vz{\vec z}

\def\ve{\vec e}\def\vk{\vec k}\def\vn{\vec n}\def\vp{\vec p}
\def\vr{\vec r}\def\vs{\vec s}\def\vt{\vec t}\def\vu{\vec u}
\def\vv{\vec v}\def\vx{\vec x}\def\vy{\vec y}\def\vz{\vec z}

\newcommand{\AdS}{\mathrm{AdS}}
\newcommand{\dd}{\mathrm{d}}
\newcommand{\eee}{\mathrm{e}}
\newcommand{\sgn}{\mathop{\mathrm{sgn}}}

\def\a{\alpha}
\def\b{\beta}
\def\g{\gamma}

\newcommand\lsim{\mathrel{\rlap{\lower4pt\hbox{\hskip1pt$\sim$}}
    \raise1pt\hbox{$<$}}}
\newcommand\gsim{\mathrel{\rlap{\lower4pt\hbox{\hskip1pt$\sim$}}
    \raise1pt\hbox{$>$}}}

\newcommand{\beq}{\begin{equation}}
\newcommand{\eeq}{\end{equation}}
\newcommand{\bea}{\begin{eqnarray}}
\newcommand{\eea}{\end{eqnarray}}
\newcommand{\bem}{\begin{pmatrix}}
\newcommand{\eem}{\end{pmatrix}}
\newcommand{\noi}{\noindent}
\newcommand{\non}{\nonumber}
\newcommand{\rdec}{\color{red}}
\newcommand{\plav}{\color{blue}}
\newcommand{\bet}{\begin{itemize}}
\newcommand{\eet}{\end{itemize}}
\newcommand{\ben}{\begin{enumerate}}
\newcommand{\een}{\end{enumerate}}


%
\bigskip

\begin{center}

{\Large\bf  Holographic thermal propagator from modularity
 }
\vspace{1cm}

\centerline{
Borut Bajc$^{a}$\footnote{borut.bajc@ijs.si} 
and 
Katarina Trailovi\'c$^{a,b}$\footnote{katarina.trailovic@ijs.si}
}
\vspace{0.5cm}
\centerline{
$^{a}$ {\it\small J.\ Stefan Institute, 1000 Ljubljana, Slovenia}
}
\centerline{
$^{b}$ {\it\small Faculty of Mathematics and Physics, University of Ljubljana,}
}
\centerline{{\it\small 1000 Ljubljana, Slovenia}}
\end{center}

\bigskip

\begin{abstract}
It is known that the holographic thermal propagator in 4 spacetime dimensions can be related to the Nekrasov-Shatashvili 
limit of the $\Omega$-deformed ${\cal N}=2$ supersymmetric $SU(2)$ Yang-Mills theory with $N_f=4$ hypermultiplets. There are two expansions 
involved: one is the expansion in small temperature which in the Seiberg-Witten language is
equivalent to the semiclassical expansion in inverse powers of the large adjoint vev and the second is the 
expansion in instanton numbers. Working in the simplified case of zero energy, we find that the latter expansion 
gives rise to quasi-modular forms which can be resummed as functions of Eisenstein series. 
The so obtained series in positive powers of small temperature shows clear signs of being asymptotic.
\end{abstract}

\clearpage

\tableofcontents

\section{Introduction}

There have been various attempts to compute the low temperature expansion of the 
holographic \cite{Maldacena:1997re,Gubser:1998bc,Witten:1998qj} thermal  
propagator of operators with conformal dimension $\Delta=\nu+2$ in 4 spacetime dimensions. 
Differently from 2 spacetime dimensions where the solution is known analytically \cite{Grozdanov:2019uhi}, 
in 4d a compact analytic form is not known. The above mentioned attempts run from the 
${\vec k}=0$ case \cite{Policastro:2001yb}, large conformal dimension limit 
\cite{Rodriguez-Gomez:2021pfh,Rodriguez-Gomez:2021mkk}, to fixed order calculations 
in the low temperature expansion \cite{Fitzpatrick:2019zqz,Parisini:2022wkb,Bajc:2022wws}. The 
method found in \cite{Fitzpatrick:2019zqz} seems the most successful among them since it gives a 
well defined and prescribed ansatz for a given term of the temperature expansion of the propagator. 
Indeed, the method has been further developed to a very efficient prescription \cite{Huang:2024wbq,Buric:2025fye}. 
Notice that these methods give only the stress-tensor sector of the OPE expansion. 
For the complete full retarded propagator one needs also the double-trace sector, which makes sure that 
at the temperature $T$ the propagator at time $\tau$ is equal to the propagator at time $1/T-\tau$, i.e. it 
satisfies the KMS relation \cite{El-Showk:2011yvt,Iliesiu:2018fao}. To get the contribution of the double-trace sector 
see for example \cite{Fitzpatrick:2019zqz,Parisini:2022wkb,Buric:2025fye,Buric:2025anb,Niarchos:2025cdg}.

Already from the first attempts it was clear that in the 5d bulk the equation to solve is the Heun equation. 
The constraint on the horizon and the conditions on the boundary need to relate the solutions 
around different singular points of the Heun equation, i.e. to solve the so called connection problem. 
How to solve it in general has been shown recently in  \cite{Bonelli:2022ten}, and used for the thermal 
propagator in \cite{Dodelson:2022yvn}. Although in this way the full propagator is in principle known, it is still written as an infinite sum, 
where higher terms need more effort to be calculated: it is essentially the calculation of the instanton 
contribution to the prepotential $F$ of a $\Omega$-deformed Nekrasov-Shatashvili \cite{Nekrasov:2009rc} limit 
of the ${\cal N}=2$ supersymmetric SU(2) gauge theory with $N_f=4$ hypermultiplets in the 
fundamental-antifundamental representation \cite{Seiberg:1994aj}, where the masses of the four quarks 
depend on the shifted scaling dimension $\nu=\Delta-2$ and the ratio of the energy of the propagator with the temperature $\omega/T$.

Schematically, the holographic retarded thermal propagator is given by \cite{Dodelson:2022yvn}
\beq
\label{GRschematic}
G_R\propto\exp{\left(- 2 \partial_\nu H(a,\nu,\omega,q)\right)},
\eeq
where the power of $q$ counts the instanton number of the instanton prepotential $H(a,\nu,\omega,q)$. 
The adjoint vev $a$ of the scalar in the vector multiplet is connected to the 3-momentum square ${\vec k}^{\,2}$ of the propagator 
through the Matone relation \cite{Matone:1995rx}

\beq
{\vec k}^{\,2}\propto q\partial_qH(a,\nu,\omega,q)+\ldots,
\eeq
where the dots represent a polynomial in $a$, $\nu$ and $\omega$.

We are interested in the propagator of the black brane and hence, we have to fix $q=\exp{(-\pi)}$ 
at the end of the calculation. This means that the propagator 
depends on the energy $\omega$, the momentum square ${\vec k}^{\,2}$ and the shifted scaling dimension $\nu$. 
The temperature dependence enters by replacing the dimensionless energy $\omega$ and momentum $\vec{k}$ with

\beq
\omega\to\omega/T\quad,\quad{\vec k}^{\,2}\to{\vec k}^{\,2}/T^2,
\eeq

\noi
i.e. as a mass unit, so that the low temperature expansion essentially means a large $\omega$ and 
large ${\vec k}^{\,2}$ expansion, and, through the Matone relation, it turns out that $a$ is large too.
Therefore, crucial to performing a low temperature expansion of the black brane propagator, is the calculation of $H$ in this limit of large ${\vec k}^{\,2}$ and $\omega$. 

\beq
H(a,\nu,\omega,q)=\sum_{n=1}^\infty q^n\frac{N_n(a,\nu,\omega)}{D_n(a)}.
\eeq

There is a well known \cite{Nekrasov:2002qd} although cumbersome prescription to get the ratios $N_n/D_n$. However, in the small temperature expansion we already encounter a problem: both $a$ {\it and} $\omega$ are large. Complications connected to it 
will be trivially bypassed by studying the limiting case of $\omega=0$:

\beq
H(a,\nu,q)=\sum_{n=1}^\infty q^n\frac{N_n(a,\nu)}{D_n(a)},
\eeq
such that we have only to expand in large $a$.
The results will thus not be general, but will nevertheless give quite some information.

The second problem is the infinite sum over powers of $q$. Since the natural expansion is in large $a$, we will 
rewrite the above sum as\footnote{Since only two quark masses are nonzero in the $\omega=0$ case, only even 
powers of $a$ appear.}

\beq
H(a,\nu,q)=\sum_{n=1}^\infty \frac{\alpha_{2n}(q,\nu)}{a^{2n}}.
\eeq
Through explicit computation and general arguments we will find that the prepotential is 
quasi-modular in $q=\exp{(i\pi\tau)}$. This property will allow for an efficient resummation in $q$ with low computational cost. Of course it remains an infinite sum over inverse powers of $a$, but since on dimensional ground $a\propto1/T$, this expansion is nothing else than the expansion we are 
looking for. We will be able to write the explicit exact form of $\alpha_{2n}(q,\nu)$ for $n$ not too large 
(we computed them up to $2n=58$ but higher coefficients could be easily found). The efficiency seems comparable to other methods 
\cite{Fitzpatrick:2019zqz,Buric:2025fye} but it is done in momentum space and in the $\omega\to0$ limit instead of the coordinate space and $\vec{k}\to0$ limit of \cite{Buric:2025fye}. This efficiency and the curious quasi-modularity of the propagator\footnote{The 
quasi-modularity of the propagator could have been guessed already from \cite{Dodelson:2022yvn} 
and is certainly well known in the ${\cal N}=2$ community.} are the main results of this paper.

The rest of this paper is organized as follows: In Section 2, we review Zamolodchikov's $q$-recursion and show its importance for the $4$-point conformal block of a $2d$ CFT as well as for the instanton part of the Nekrasov partition function. In Section 3, we introduce the retarded two-point function of a black brane, which we bring to a compact form for the limit of a large adjoint scalar vev $a$. This limit corresponds, due to the Matone relation, to the low-temperature limit in which we are interested. 
In Section 4, we explain the $S$-duality constraints on the prepotential and elaborate on how this leads to quasi-modular forms. We then use this quasi-modularity of the prepotential coefficients in the large $a$-expansion to calculate them efficiently by matching with Zamolodchikov's $q$-recursion. 
Finally, in Section 5, we give the low temperature expansion of the propagator. In Appendix A, we review the holographic two-point functions for the black brane. In Appendix B, we give the Eisenstein series and some of their properties. In Appendix C, we present the formulas of ${\alpha}_{2n}$ up to $2n=18$, while in Appendix D, we give the explicit expansion of the propagator  up to $T^{40}$.

\section{Zamolodchikov's $q$-recursion}

In this section we rewrite the instantonic part of the Nekrasov-Shatashvili prepotential in terms of the 
renormalised instanton number $q$ instead of the unrenormalised instanton number $t$ used in \cite{Dodelson:2022yvn} applying the AGT corrspondence \cite{Alday:2009aq} and the Zamolodchnikov solution \cite{Zamolodchikov:1984eqp,Zamolodchikov:1987avt,Zamolodchikov:1995aa,Poghossian:2009mk}. This will then be used in the next section.

In general, the 4-point conformal block of a $2d$ CFT can be written as 
\beq
\label{4ptcb}
\begin{split}
\mathcal{F}(\Delta,\Delta_i,t)=&(16 q)^{-\alpha^2}t^{Q^2/4-\Delta_1-\Delta_2}(1-t)^{Q^2/4-\Delta_1-\Delta_3}\\
&\times \theta_3(q)^{3Q^2-4(\Delta_1+\Delta_2+\Delta_3+\Delta_4)}\mathcal{H}(q,\Delta,\vec{\mu}),
    \end{split}
\eeq
where $\Delta_{i=1,2,3,4}$ are the conformal dimensions of the external primary fields at the positions $0,t,1,\infty$, \ $\Delta$ is the internal (intermediate) conformal dimension and $Q$ is the background charge, related to the central charge of the Virasoro algebra via $c=1-6 Q^2$.
Additionally, we can parametrize the dimensions as 
\begin{equation}
    \Delta_i=\frac{Q^2}{4}-\lambda_i^2, \quad \Delta=\frac{Q^2}{4}-\alpha^2,
\end{equation}
where $\alpha$ is the internal momentum that appears in \eqref{4ptcb} and $\lambda_i$ gives the parameters $\vec{\mu}$ inside $\mathcal{H}$:
\begin{equation}
\mu_1=\lambda_1+\lambda_2+\frac{Q}{2}\ , \quad \mu_2=\lambda_1-\lambda_2+\frac{Q}{2}\ ,\quad
  \mu_3=\lambda_3+\lambda_4+\frac{Q}{2}\ ,\quad \mu_4=\lambda_3-\lambda_4+\frac{Q}{2}\ .  
\end{equation}
In the AGT correspondence \cite{Alday:2009aq}, the central charge of the Liouville theory is related to the Nekrasov deformation parameters $\epsilon_1, \epsilon_2$ of the $\Omega$-background in the $4d$ gauge theory as
\begin{equation}
 Q=\frac{\epsilon_1+\epsilon_2}{\sqrt{\epsilon_1 \epsilon_2}}
\end{equation}
and the coupling $t$ is related to the renormalized coupling $q$ via
\beq
q(t)=\exp{\left(-\pi\frac{K(1-t)}{K(t)}\right)}\quad,\quad K(t)=\frac{1}{2}\int_0^1\frac{dx}{\sqrt{x(1-x)(1-tx)}},
\eeq

\noi
i.e.\ conversely 

\beq
\label{t}
t(q)=\left(\frac{\theta_2(q)}{\theta_3(q)}\right)^4, \quad \text{ where } \quad  \theta_2(q)=\sum_{n=-\infty}^\infty q^{(n+1/2)^2}, \quad \theta_3(q)=\sum_{n=-\infty}^\infty q^{n^2}.
\eeq
In order to compute $\mathcal{H}$, we can use the Zamolodchikov $q$-recursion \cite{Zamolodchikov:1984eqp,Zamolodchikov:1987avt,Zamolodchikov:1995aa}:
\beq
\label{H}
{\cal H}(q,\Delta,\vec{\mu})=1+\sum_{m,n=1}^\infty \frac{q^{mn}R_{m,n}(\vec{\mu})}{\Delta-\Delta_{m,n}}{\cal H}(q,\Delta_{m,n}+mn,\vec{\mu}),
\eeq

\noi
where

\beq
\Delta_{m,n}=\lambda_{1,1}^2-\lambda_{m,n}^2\quad,\quad 
\lambda_{m,n}=\frac{m\epsilon_1+n \epsilon_2}{2\sqrt{\epsilon_1\epsilon_2}}
\eeq
and

\beq
\label{R}
R_{m,n}(\vec{\mu})=\frac{2\prod_{r,s}\prod_{i=1}^4\left(\mu_i-\frac{Q}{2}-\lambda_{r,s}\right)}{\prod_{k,l}'\lambda_{k,l}}
\eeq

\noi
with

\beq
r=-m+1,-m+3,\ldots,m-1\non
\eeq
\beq
s=-n+1,-n+3,\ldots,n-1\non
\eeq
\beq
k=-m+1,-m+2,\ldots,m-1,m\non
\eeq
\beq
l=-n+1,-n+2,\ldots,n-1,n\non
\eeq

\noi
and the prime over the product in the denominator denotes that we have 
to skip the pairs $(k,l)=(m,n)$ and $(0,0)$.

We are interested in the $\mathcal{N}=2$ SYM theory with $SU(2)$ gauge group and extra four anti-fundamental hypermultiplets, $f=4$. For this case, the instanton part of the Nekrasov partition function \cite{Nekrasov:2002qd} , after applying the AGT conjecture, is given by \cite{Alday:2009aq}
\beq
Z_\text{inst}^{(4)}(a,m_i,t,\epsilon_1,\epsilon_2)=t^{\Delta_1+\Delta_2-\Delta} (1-t)^{2(\lambda_1+Q/2)(\lambda_3+Q/2)}\mathcal{F}(\Delta,\Delta_i,t)
\eeq
with $a=\alpha\sqrt{\epsilon_1 \epsilon_2}$ being the vev of the adjoint scalar in the vector multiplet, parameterizing the Coloumb branch and $m_i=(\mu_i-Q/2)\sqrt{\epsilon_1 \epsilon_2}$ being the masses of the four hypermultiplets, which using $\vec{a}=(a_t, a_0, a_1, a_\infty)=(\lambda_1,\lambda_2,\lambda_3,\lambda_4)\sqrt{\epsilon_1 \epsilon_2}$, are given by
\beq
\label{masses}
m_1=a_t+a_0\quad,\quad
m_2=a_t-a_0\quad,\quad
m_3=a_1+a_\infty\quad,\quad
m_4=a_1-a_\infty.
\eeq
The instanton part of the Nekrasov partition function, after applying \eqref{4ptcb}, reduces to 
\beq
\label{NSpartZam}
Z_\text{inst}^{(4)}(a,m_i,q,\epsilon_1,\epsilon_2)=\left(\frac{t(q)}{16 q}\right)^{\alpha^2} (1-t(q))^{\frac{1}{4}(Q-\sum_{i=1}^4 \mu_i)^2}\theta_3(q)^{2\sum_{i=1}^4 \mu_i(\mu_i-Q)+Q^2} \mathcal{H}(q,\Delta,\vec{\mu})
\eeq
and hence the instanton part of the Nerkrasov partition function in the $\mathcal{N}=2$ SYM, $N_c=2, N_f=4$ theory can be computed with the help of Zamolodchikov's $q$-recursion.

While the instanton part of the Seiberg-Witten prepotential \cite{Seiberg:1994aj} is given by the limit
\beq
F_\text{inst}^\text{SW}=-\lim_{\epsilon_1,\epsilon_2 \rightarrow 0} \epsilon_1 \epsilon_2 \log Z_\text{inst}^{(4)},
\eeq
taking the Nekrasov-Shatashvili (NS) limit $\epsilon_1=1, \epsilon_2\rightarrow 0$ \cite{Nekrasov:2009rc} gives the instanton part of the NS function 
\beq
\label{instantonNS}
F_\text{inst}^\text{NS}=\lim_{\epsilon_1\rightarrow 1,\epsilon_2 \rightarrow 0} \epsilon_1 \epsilon_2 \log[(1-t)^{-2(Q/2+\lambda_1)(Q/2+\lambda_3)} Z_\text{inst}^{(4)}\ ],
\eeq
where the $(1-t)^{-2(Q/2+\lambda_1)(Q/2+\lambda_3)}$ prefactor is present in order to remove the $U(1)$ factor from the $U(2)$ Nekrasov partition function \cite{Alday:2009aq}.

We can write the NS free energy in a more explicit way using \eqref{NSpartZam} and \eqref{instantonNS}, which will be useful later on\footnote{From now on the quantity $F_{\text{inst}}^{\text{NS}}$ will be called simply $F$.},
\bea
\label{F}
F(q,a,\vec{a})&=&a^2\left(\log{\left(t(q)\right)}-\log{\left(16q\right)}\right)
+\left(a_t^2+a_1^2-\frac{1}{4}\right)\log{(1-t(q))}\non\\
&&+\left(a_0^2+a_t^2+a_1^2+a_\infty^2-\frac{1}{4}\right)\log{\theta_3^4(q)}
+H(q,a,\vec{a})
\eea

\noi
where we defined

\beq
H(q,a,\vec{a})\equiv\lim_{\epsilon_1\to1,\epsilon_2\to0}\epsilon_1\epsilon_2\log{{\cal H}(q,\Delta,\vec{\mu})}. \label{prepot}
\eeq

\noi
Therefore, we can compute the NS free energy using the Zamolodchikov $q$-recursion.

\section{The retarded Green's function of the black brane}

We study a holographic conformal field theory defined on 
$S^1 \times \mathbb{R}^3$, which is dual to a black brane geometry. 
The two-point function of a scalar operator in this CFT is obtained by 
solving the bulk wave equation for its dual scalar field in the black brane 
background, and then extracting the retarded Green’s function.  

For a scalar operator of conformal dimension $\Delta = 2 + \nu$, 
the retarded Green’s function is given by the ratio of the response to the 
source at the AdS boundary. This quantity can be computed explicitly by 
solving the associated Heun equation, characterized by singular points at 
$z = 0, t, 1, \infty$ with parameters $a_0, a_t, a_1, a_\infty$, and $u$\cite{Dodelson:2022yvn}:
\beq
\label{heunwiththetas}
\left[ \frac{d^2}{dz^2} + \frac{\frac{1}{4} - a_0^2}{z^2} + \frac{\frac{1}{4} - a_1^2}{(z-1)^2} + 
\frac{\frac{1}{4} - a_t^2}{(z-t)^2} + \frac{a_0^2 + a_1^2 + a_t^2 - a_\infty^2 - \frac{1}{2}}{z(z-1)} 
- \frac{(1-t)u}{z(z-1)(z-t)} \right] \chi = 0,
\eeq(for more details see Appendix \ref{BHBB}). The Heun parameters 
are related to physical quantities of the Minkowski black brane as 
\beq
\label{BBpar}
t=\frac{1}{2}\;,\;\vec{a}=(a_0,a_t,a_1,a_\infty)=
\left(0,\frac{i\omega}{4\pi},\frac{\nu}{2},\frac{\omega}{4\pi}\right)\;,\;
u=\frac{\omega^2-2k^2}{8\pi^2}-\frac{\nu^2}{4}.
\eeq
It has been shown \cite{Bonelli:2022ten} that the exact solution to the connection problem of the Heun equation 
can be expressed through the Nekrasov-Shatashvili prepotential \cite{Nekrasov:2009rc} of 
the $\mathcal{N}=2$ supersymmetric $SU(2)$ Yang-Mills theory with $N_f=4$ hypermultiplets, where 
the masses of the hypermultiplets are combinations of $a_0,a_t,a_1,a_\infty$ (see \eqref{masses}), $t$ is the instanton counting parameter and $u$ parametrizes the moduli space of vacua. The latter is connected with the parameter $a$ corresponding to the vev of the scalar in the vector multiplet via the Matone relation \cite{Matone:1995rx}
\beq
\label{matone}
u=-a^2+a_t^2+a_0^2-\frac{1}{4}+t\partial_tF,
\eeq
where $F$ is the instanton part of the NS free energy defined in \eqref{F}.

We are interested in the limit where $\omega=0$ and $|\vec{k}|\rightarrow \infty$,\ i.e. such that when promoted to dimensionful parameters $\omega\rightarrow \omega/T$ and $|\vec{k}|\rightarrow |\vec{k}|/T$, we are in the zero energy and low temperature limit (low with respect to the three-momentum). In this case, we see that the Matone relation implies $a \rightarrow +\infty$.

Following \cite{Dodelson:2022yvn}, we can write the propagator as
(modulo exponentially small terms for the limit we are interested in, $a\to+\infty$)

\beq
\label{G}
G_R=\pi^{4a_1}e^{-\partial_{a_1}F}\frac{M(a,a_1;a_\infty)}{M(a,-a_1;a_\infty)},
\eeq

\noi
where

\beq
M(\alpha_0,\alpha_1;\alpha_2)=\frac{\Gamma(-2\alpha_1)\Gamma(1-2\alpha_0)}
{\Gamma(1/2-\alpha_0-\alpha_1+\alpha_2)\Gamma(1/2-\alpha_0-\alpha_1-\alpha_2)}.
\eeq
Now, we want to perform a large $a$ expansion of this formula and finally bring the propagator into a compact and useful form for the low temperature expansion.  In order to do so, we first need to make use of the gamma function formula in the large $a$ limit, given by  \cite{neven}
\beq
\log{\Gamma(a+\alpha)}=(a+\alpha-1/2)\log{a}-a+\frac{1}{2}\log{(2\pi)}+\sum_{n=1}^\infty
\frac{(-1)^{n+1}B_{n+1}(\alpha)}{n(n+1)a^n}
\eeq

\noi
with $B_n(\alpha)$ the Bernoulli polynomials, so that (an alternative is to use \cite{Gamma})

\beq
\frac{\Gamma(a+\alpha)}{\Gamma(a+\beta)}=a^{\alpha-\beta}
\exp{\left(\sum_{n=1}^\infty\frac{(-1)^{n+1}}{n(n+1)a^n}\left(B_{n+1}(\alpha)-B_{n+1}(\beta)\right)\right)}.
\eeq

%
%
%
%
\noi
Note that in the case of $\alpha+\beta=1$, we have $B_{l}(1-\alpha)=(-1)^lB_{l}(\alpha)$ for any $l>1$.

%

Then, we rewrite the Zamolodchikov $q$-recursion into an explicit form
\beq
\label{explicitZamolodchikov}
\mathcal{H}(q,\Delta,\vec{\mu})=1+\sum_{\chi=1}^\infty c_\chi(a,\vec{\mu},\epsilon_j) q^{\chi},
\eeq
where
\beq
c_\chi(a,\vec{\mu},\epsilon_j)=\sum_{k=1}^\chi \sum_{\sum_{i=1}^k m_i n_i=\chi} \frac{\epsilon_1 \epsilon_2}{\left(-a^2+\left(\frac{m_1\epsilon_1+n_1 \epsilon_2}{2}\right)^2\right)} \frac{\prod_{i=1}^k R_{m_i n_i}(\vec{\mu})}{\prod_{i=1}^{k-1} \Delta_{m_i n_i}+m_i n_i -\Delta_{m_{i+1} n_{i+1}}}.
\eeq
We use this formula to expand \eqref{prepot} for large $a$ and obtain 
\beq
\label{H}
H(a,q,\nu)=\sum_{n=1}^\infty \frac{\alpha_{2n}(q,\nu)}{a^{2n}}.
\eeq
Now, with all of the steps before and using \eqref{F}, we can finally write
the propagator in a compact form:

%
\beq
\label{GRzGR0}
\boxed{\frac{G_R}{G_{R}^{0}}=\left(\frac{a^2}{x^2}\right)^\nu
\exp{\left(-2\sum_{n=1}^\infty\frac{\partial_\nu\tilde\alpha_{2n}}{a^{2n}}\right)}}
\eeq

\noi
with the conformal propagator being

\beq
G_{R}^{0}=\frac{\Gamma(-\nu)}{\Gamma(\nu)}\left(\frac{k}{2}\right)^{2\nu}.
\eeq
To obtain this formula, we made additionally use of the following:

First we rewrote
\beq
\label{tildealpha}
\frac{B_{2n+1}\left(\frac{1+\nu}{2}\right)}{n(2n+1)}
+\partial_\nu\alpha_{2n}=\partial_\nu\tilde\alpha_{2n}
\eeq
Notice that $\alpha_{2n}$ is a polynomial in $q^2$ with no constant term, since it is a pure instanton contribution, while 
$\tilde\alpha_{2n}$ has on top of that also a $q$-independent piece. This rewriting will be useful since $\tilde\alpha_{2n}$ will have a simple and computationally useful form, as we will see later.

Second, we defined 
\beq
x^2\equiv \frac{k^2}{8\pi\Gamma(3/4)^4}
\eeq
and then we used that the connection between $a$ and $k$ (resp. $x$) is the Matone relation

\beq
\label{matone}
\boxed{x^2=a^2-\frac{\nu^2-1}{4\pi}-\sum_{n=1}^\infty\frac{q\partial_q\tilde\alpha_{2n}}{a^{2n}}},
\eeq
which by inversion gives

\beq
\label{aofx}
\boxed{a^2=x^2+\sum_{n=0}^\infty\frac{\beta_{2n}}{x^{2n}}.}
\eeq

\noi
To compute $\beta_{2n}$ one has to plug in \eqref{aofx} into \eqref{matone} and expand in large $x$. We give few lower terms: 
\beq
\beta_0=\frac{\nu^2-1}{4\pi}\quad,\quad
\beta_2=q\partial_q\tilde{\alpha}_2\quad,\quad
\beta_4=q\partial_q\tilde{\alpha}_4-\frac{\nu^2-1}{4\pi}q\partial_q\tilde{\alpha}_2.
\eeq
Note that although we keep the dependence on $q$ (or equivalently $t$) general in the intermediate steps because of derivative terms, in the final formula \eqref{GRzGR0} we set $t=1/2$, i.e.\ $q=\exp(-\pi)$.

Our goal is to obtain the low temperature expansion of the propagator, therefore the procedure consists of first calculating $\tilde\alpha_{2n}(q,\nu)$, then finding explicitly (\ref{aofx}), i.e. $a(k)$ through inversion of (\ref{matone}), 
plugging everything in (\ref{GRzGR0}) and after replacing $|\vec{k}|\to |\vec{k}|/T$, expanding the propagator in positive powers of $T$.

\section{$\mathcal{S}$-duality constraints on the prepotential: quasi-modular forms} 
In this section, we analyze the emergence of quasi-modular forms in the Nekrasov-Shatashvili limit of the $\Omega$-deformed $\mathcal{N}=2$, $\mathrm{SU}(2)$ gauge theory with $N_f = 4$, arising from the modular properties of the Seiberg-Witten geometry and its quantum deformation \cite{Minahan:1997if,Billo:2013jba,Billo:2015pjb}.

The low-energy effective action on the Coulomb branch of the gauge theory, is determined by the prepotential $F_{\text{NS}}(a)$, which can be written as \cite{Dodelson:2022yvn}

\beq
F_{\text{NS}}=-a^2\log{(t)}+F_{1-\text{loop}}+F,
\eeq
with $F_{1-\text{loop}}$ being the perturbative contribution up to 1-loop and $F$ the instanton contribution.

The 1-loop contribution  $F_{1-\text{loop}}$ can be expanded in large $a$ as \cite{Billo:2013fi}

\beq
\label{F1loop}
F_{1-\text{loop}}=a^2\log{(16)}-\frac{\nu^2-1}{4}\log{\left(\frac{a^2}{\Lambda^2}\right)}
+\sum_{n=1}^\infty\frac{f_{2n}}{a^{2n}},
\eeq

\noi
while the large $a$-expansion of the instanton contribution, using eqs. (\ref{F}), (\ref{t}) and (\ref{H}), is given by

\beq
F=a^2\log{(t)}-a^2\log{(16q)}+\frac{\nu^2-1}{4}\log{({\theta_4}^4)}
+\sum_{n=1}^\infty\frac{\alpha_{2n}}{a^{2n}}.
\eeq

\noi
The full Nekrasov-Shatashvili prepotential then becomes

\beq
\label{FNS}
F_{\text{NS}}=-i\pi\tau a^2+F_q
\eeq

\noi
where $\tau$ is the gauge coupling in the IR,

\beq
\tau=\frac{\theta_\text{IR}}{\pi}+i\frac{8\pi}{g_\text{IR}^2}\ ,
\eeq

\noi
which is related to the instanton counting parameter as

\beq
q=e^{i\pi\tau},
\eeq

\noi
and  

\beq
\label{calF}
F_q=-\frac{\nu^2-1}{4}\log{\left(\frac{a^2}{\Lambda^2{\theta_4}^4}\right)}
+\sum_{n=1}^\infty\frac{\alpha_{2n}+f_{2n}}{a^{2n}},
\eeq
with an arbitrary scale $\Lambda$.

%
%
%

The same theory can be also described using the dual variable
\beq
\label{duala}
a_D = \frac{1}{2\pi i} \frac{\partial F_\text{NS}}{\partial a}.
\eeq

Now, $\tau$ transforms under the modular group, reflecting the electric-magnetic duality symmetries of the Seiberg-Witten, i.e.\ it transforms under the generators $\mathcal{S}$ and $\mathcal{T}$ of the modular group $PSL(2,\mathbb{Z})$ as 
\beq
\mathcal{S}: \tau \rightarrow -\frac{1}{\tau}, \quad \mathcal{T}: \tau\rightarrow \tau+1
\eeq
while $a$ and $a_D$ transform as 
\beq
\label{STtrafo}
\mathcal{S}: a \rightarrow a_D, \ a_D \rightarrow -a, \quad \mathcal{T}: a \rightarrow a, \ a_D \rightarrow a_D+a.
\eeq
The transformation $\mathcal{T}$ on the instaton part of the prepotential $F_\text{inst}(a)$ is trivial, since $F_\text{inst}(a)=\sum_k q^{2k} F_k(a)$ and $q\rightarrow-q,\ a\rightarrow a$. So the prepotential transforms as \cite{Billo:2015pjb}

\beq
{\cal T}\quad:\quad F_\text{NS}(a)\to F_\text{NS}(a)-i\pi a^2
\eeq

\noi
providing $\Lambda^2{\theta_4}^4$ remains invariant under ${\cal T}$.

However, the $\mathcal{S}$-transform is not trivial, it maps the theory with $a$ to the dual theory with $a_D$ and therefore it should map the prepotential to its Legendre transform 
\beq
\label{Legendre}
\mathcal{S}: F_\text{NS}(a)\rightarrow F_\text{NS}(a)- 2\pi i a \hspace{0.03cm} a_D=F_{\text{NS}}(a)-a\frac{\partial F_\text{NS}}{\partial a}.
\eeq
This $\mathcal{S}$-duality condition gives a strong constraint on the prepotential and leads to the fact that the prepotential can be written in terms of quasi-modular forms.

To see the modularity directly, let us compute $\alpha_{2n}(q,\nu)$ from \eqref{H} using \eqref{explicitZamolodchikov}.
The explicit formula for $\alpha_{2n}$, which comes from the large $a$ expansion of $H$, is given by
\begin{equation}
\label{alphaformula}
\begin{split}
\alpha_{2n}&=- \lim_{\substack{\epsilon_1 \to 1 \\ \epsilon_2 \to 0}} \left[\sum_{m=1}^n \frac{(\epsilon_1\epsilon_2)^{m+1}}{m} \sum_{\substack{\sum_{l=0}^\infty k_l=m \\ \sum_{l=0}^\infty k_l (l+1)=n}}\frac{m!}{\prod_{l=0}^\infty k_l !}\prod_{l=0}^\infty \left(\sum_{\chi=1}^\infty \tilde{c}_\chi^l q^\chi \right)^{k_l}\right]
\end{split}
\end{equation}
with 
\beq
\tilde{c}_\chi^l=\sum_{k=1}^\chi \sum_{\sum_{i=1}^k m_i n_i=\chi}\left(\frac{m_1 \epsilon_1+n_1 \epsilon_2}{2}\right)^{2l} \frac{\prod_{i=1}^k R_{m_i n_i}}{\prod_{i=1}^{k-1} \Delta_{m_i n_i}+m_i n_i -\Delta_{m_{i+1} n_{i+1}}}.
 \eeq
The first few $\alpha_{2n}$'s are  
\beq
\begin{split}
\alpha_2=&-\frac{( \nu^2-1)^2}{4} q^2 (1 + 3 q^2 + 4 q^4 + 7 q^6 + 6 q^8 + 12 q^{10} + 
   8 q^{12}+\ldots),\\
\alpha_4=&
-\frac{(\nu^2-1)^2}{8} q^2  \left(2 + 3 q^2 (7 - \nu^2) + 
   8 q^4 (9 - 2 \nu^2) + 
   q^6 (191 - 45 \nu^2)\right. \\ & \left.+ 
   12 q^8 (29 - 8 \nu^2) + 36 q^{10} (19 - 5 \nu^2)+ 16 q^{12} (61 - 18 \nu^2)+\ldots \right)\\
   \alpha_6=&-\frac{(\nu^2-1)^2}{128} q^2  \left( 32 + 
   q^2 (1301 - 250 \nu^2 + 5 \nu^4) + 
   32 q^4 (329 - 90 \nu^2 + 5 \nu^4) \right. \\ &+ 
   q^6 (47159 - 14430 \nu^2 + 1095 \nu^4) 
   +64 q^8 (2254 - 758 \nu^2 + 67 \nu^4) \\ & \left.+  
   12 q^{10} (31227 - 10790 \nu^2 + 1035 \nu^4) + 
   64 q^{12} (12469 - 4530 \nu^2 + 465 \nu^4)+\ldots  
    \right).
   \end{split}
   \raisetag{80pt} 
\eeq
Note, because $\omega=0$, it follows that $R_{m n}=0$ for $m$ and $n$ both odd, which implies $\tilde{c}^l_\chi=0$ for $\chi$ odd and hence only even powers of $q$ appear.

Now, we observe that these coefficients together with the 1-loop constant term are quasi-modular forms 
\cite{zagier}, i.e.
\beq
\label{alphamod}
\alpha_{2n}(q,\nu)+f_{2n}(\nu)=\sum_{2k+4l+6m=2n}c_{k,l,m}(\nu){E_2}^k(q){E_4}^l(q){E_6}^m(q)\equiv 
\tilde{\alpha}_{2n}(q,\nu)
\eeq
where $E_{2,4,6}(q)$ are the Eisenstein series summarised in Appendix \ref{eisenstein}, and that the constant term is given by
\beq
f_{2n}(\nu)=\frac{B_{2n+2}\left(\frac{1+\nu}{2}\right)-B_{2n+2}}{(2n+1)n (n+1)},
\eeq
where $B_{2n+2}=B_{2n+2}(0)$ are the Bernoulli numbers\footnote{Note also that since $q=0$ gives $\mathcal{H}=1$, it follows $\alpha_{2n}=0$ and due to $E_{2,4,6}(q=0)=1$ we have $\sum_{2k+4l+6m=2n} c_{k,l,m}=f_{2n}$.}. 
Therefore, the definition of $\tilde{\alpha}_{2n}$ in \eqref{alphamod} matches the condition in \eqref{tildealpha} and $\tilde{\alpha}_{2n}$ is a quasi-modular form of weight $2n$.

So (\ref{calF}) can be written as

\beq
F_q=-\frac{h_0}{2}\log{\left(\frac{a^2}{\Lambda^2{\theta_4}^4}\right)}
+\tilde H(q,a,\nu)
\eeq

\noi
with

\beq
h_0=\frac{\nu^2-1}{2}
\eeq

\noi
and

\beq
\tilde H(q,a,\nu)=\sum_{n=1}^\infty\frac{\tilde\alpha_{2n}(q,\nu)}{a^{2n}},
\eeq

\noi
i.e. with exactly the same parameters needed for the 
propagator (\ref{GRzGR0}). 

The quasi-modularity of $\tilde{\alpha}_{2n}$ comes from the modular anomaly: if $\mathcal{S}: a^2\rightarrow \tau^2 a^2 $, i.e.\ $a^2$ would transform as a modular form of weight $2$, the instanton contribution (minus the constant piece eq. 
(\ref{alphamod})) would read

\beq
\label{alphamodular}
\tilde\alpha_{2n}(q,\nu)\sim\sum_{4l+6m=2n}c_{0,l,m}(\nu) E_4^l(q) E_6^m(q).
\eeq
since any modular form of positive weight can be written with $E_4$ and $E_6$ only \cite{zagier}. 
However, 
the above transformation for $a$ is only classical, 
the correct one is given in eq. \eqref{STtrafo} together with \eqref{duala} and therefore the approximate (\ref{alphamodular}) 
gets replaced by the now exact quasi-modular $\tilde{\alpha}_{2n}$ in \eqref{alphamod}.

Let's now summarise \cite{Billo:2013jba, Billo:2015pjb} to show that 
${\cal S}$ duality fixes uniquely the quasi-modular coefficients $c_{k,l,m}$ for $k>0$ from the knowledge of the modular coefficients $c_{0,l,m}$. On the one side, we have

\beq
{\cal S}\left(F_{\text{NS}}(\tau,a)\right)=F_{\text{NS}}(-1/\tau,a_D),
\eeq

\noi 
while on the other this is just the Legendre transform (\ref{Legendre}). 
Equating them and using (\ref{FNS}) brings us to

\beq
\label{calF2}
F_q(-1/\tau,a_D)=F_q(\tau,a)-\frac{1}{4\pi i\tau}\left(\frac{\partial F_q(\tau,a)}{\partial a}\right)^2.
\eeq

We can rewrite the dual variable \eqref{duala} as 

\beq
a_D=\tau\left(-a+\frac{\delta}{12a}\left(-h_0+a\frac{\partial \tilde{H}}{\partial a}\right)\right)
\eeq

\noi
where we introduced 

\beq
\delta=\frac{6}{i\pi\tau}.
\eeq

\noi
Now, we want to analyze how the operation 

\beq
\tau\to-1/\tau\quad,\quad a\to a_D
\eeq

\noi 
acts on $\tilde\alpha_{2n}/a^{2n}$. Due to (\ref{alphamod}) and 
the behaviour of $E_{2k}$ (see appendix \ref{eisenstein}) 
this operation is equivalent to 

\beq
E_2\to E_2+\delta\quad,\quad a\to 
a-\frac{\delta}{12a}\left(-h_0+a\frac{\partial \tilde{H}}{\partial a}\right).
\eeq
Taking now that $\Lambda^2\theta_4^4$ transforms as a modular form of weight 2 we can write 
(\ref{calF2}) with a little abuse of notation as
\bea
&&\tilde H
\left(E_2+\delta,a-\frac{\delta}{12a}\left(-h_0+a\frac{\partial{\tilde H}}{\partial a}\right)\right)-\tilde H(E_2,a)\non\\
&=&h_0\log{\left(1-\frac{\delta}{12a^2}\left(-h_0+a\frac{\partial\tilde H}{\partial a}\right)\right)}
-\frac{\delta}{24a^2}\left(-h_0+a\frac{\partial\tilde H}{\partial a}\right)^2.
\eea
This equation is satisfied for arbitrary $\delta$ if 

\beq
\frac{\partial\tilde H}{\partial E_2}=\frac{1}{24a^2}\left(-h_0+a\frac{\partial \tilde H}{\partial a}\right)^2.
\eeq
Expanding $\tilde H$ with inverse powers of $a^2$, gives

\beq
\label{lje1}
\frac{\partial\tilde\alpha_2}{\partial E_2}=\frac{h_0^2}{24}
\eeq

\noi
and for any integer $l>1$\footnote{This coincides with  \cite{Billo:2013fi}, providing for $l>0$
$h_l=l2^{l+1}\tilde\alpha_{2l}$, the masses of the hypermultiplets are taken as $(\nu/2,\nu/2,0,0)$, 
while the parameters of the $\Omega$-deformation are taken $\epsilon_1=1$ and $\epsilon_2=0$ 
(Nekrasov-Shatashvili limit).}

\beq
\label{lvecjiod1}
\frac{\partial\tilde\alpha_{2l}}{\partial E_2}=\frac{l-1}{6}h_0\tilde \alpha_{2(l-1)}
+\frac{1}{6}\sum_{i=1}^{l-2}i(l-i-1)\tilde\alpha_{2i}\tilde\alpha_{2l-2i-2}.
\eeq
The procedure to compute $\tilde{\alpha}_{2n}$ (or $\alpha_{2n}$) is therefore very simple: It is enough to compute a finite number of powers of $q$ for $\tilde\alpha_{2n}$ via \eqref{alphaformula}  to 
determine the finite number of coefficients $c_{k,l,m}(\nu)$ using \eqref{lvecjiod1} for the coefficients with $k\neq 0$ and matching \eqref{alphamod} to \eqref{alphaformula} for the coefficients with $k=0$.\footnote{The number of powers for $q$ that one needs to evaluate with this procedure is very small and hence the computation of $\tilde\alpha_{2n}$ is relatively fast. For example, up to $2n=58$ it is enough to compute only up to $q^8$ (or even less). }

The first few terms are
\beq
\label{alphtil8}
\begin{split}
\tilde\alpha_2=& \frac{\left(\nu ^2-1\right)^2}{96} (E_2) \\
\tilde\alpha_4=&\frac{\left(\nu ^2-1\right)^2}{11520} \left(\left(5 E_2^2+E_4\right) \nu ^2\right. \\
&\left.+\left(-5 E_2^2-13 E_4\right)\right)\\
\tilde\alpha_6=& \frac{\left(\nu ^2-1\right)^2}{5806080} \left(\left(175 E_2^3+84 E_4 E_2+11 E_6\right) \nu ^4\right.\\
&-2 \left(175 E_2^3+588 E_4 E_2+227 E_6\right) \nu ^2\\
&+\left.\left(175 E_2^3+1092 E_2 E_4+3323 E_6\right)\right)\\
\tilde\alpha_8=& \frac{\left(\nu ^2-1\right)^2}{92897280}
 \left(\left(245 E_2^4+196 E_4 E_2^2+44 E_6 E_2+19 E_4^2\right) \nu ^6\right.\\
 &-3 \left(245 E_2^4+980 E_4 E_2^2+620 E_6 E_2+339 E_4^2\right) \nu ^4\\
 &+3 \left(245 E_2^4+1764 E_4 E_2^2+5036 E_6 E_2+4883 E_4^2\right) \nu ^2\\
 &+\left.\left(-245 E_2^4-2548 E_2^2 E_4-13292 E_2 E_6-62035 E_4^2\right)\right)
 \end{split}
 \eeq
and up to $2n=18$ can be found in Appendix \ref{appendixalphatilde}. While we computed up to $2n=58$ in our analysis, more terms could be easily calculated. The general form is 

\beq
\tilde\alpha_{2n}=(\nu^2-1)^2\sum_{2k+4l+6m=2n}E_2^{k}E_4^{l}E_6^{m}\sum_{s=0}^{n-1}c_{k,l,m,s}\nu^{2s}.
\eeq
For example, one finds that

\beq
c_{3,10,2,16}=\frac{8542756834073873188278770787992128999705564764412164103052703057689}
{136719191143756950712975380634504245962798491332367173550080000}
\eeq
Although this coefficient will not enter the final expression for the propagator, since $E_6(q=\exp{(-\pi)})=0$, 
it nevertheless shows a generic pattern: it is large. And, what makes it particular interesting, one finds that such 
numbers increase with $n$. For example, the numerical coefficients in front of $E_4^n\nu^2$ of $\tilde\alpha_{4n}$ are

\bea
c_{0,n,0,1}&=&\left(0.0000868056,0.00015769,0.000775588,0.0131216,0.581048,57.3043,\right.\non\\
&&\left.11013.7,3.76087\times 10^6,2.12402\times 10^9,1.87667\times 10^{12},2.48054\times 10^{15},\right.\non\\
&&\left.4.72847\times 10^{18},1.26077\times 10^{22},4.58196\times 10^{25},\ldots\right)
\eea

\noi
or, the coefficients in front of $E_4^n\nu^4$ of $\tilde\alpha_{4n}$ are

\bea
c_{0,n,0,2}&=&\left(0.,-0.0000109476,-0.0000927139,-0.00186056,-0.0899762,-9.34756,\right.\non\\
&&\left.-1860.93,-651716.,-3.7515\times 10^8,-3.36431\times 10^{11},-4.50028\times 10^{14},\right.\non\\
&&\left.-8.66299\times 10^{17},-2.32881\times 10^{21},-8.52239\times 10^{24},\ldots\right).
\eea

\section{Low temperature expansion of the black brane propagator}
In this section, we finally want to compute the low temperature expansion of the black brane propagator, using the results from the previous sections.

First, note that the odd powers of $T$ are always zero, since the dependence on $k$ in the Matone relation is always quadratic. Second, we explicitly checked that the coefficients of $T^{4n-2}$ vanish in all the cases considered. This result is correct on general grounds\footnote{To see it one must either remember that 
the contributions we are computing come for multi-stress-tensors, or see the Heun equation which explicitly depend only on $T^4$.}, but is not obvious from the computational point of view, since all even coefficients appear in the instanton expansion of $H$. Also, the propagator as a function of $a^2$, i.e.\ \eqref{GRzGR0}  before using \eqref{matone}, has all terms proportional to $1/a^{2n}$ with $n\geq 3$ non-zero. Thus, there is a cancellation at work.

However, there is not even a partial cancellation for the terms proportional to $T^{4n}$: the large magnitude of $\tilde\alpha_{2n}$ causes the 
expansion of the propagator in powers of $\hat T\equiv  (T/k)^4$ to be asymptotic. 
As an example, the expansion of the $\nu^{5}$ term of the propagator is

\beq
\begin{split}
\left.\frac{G_R}{G_{R}^0}\right|_{\nu^{5}}=&-12.9879 \hat{T}
- 68317.4 \hat{T}^2
- 1.43349 \times 10^9 \hat{T}^3
- 1.10667 \times 10^{14} \hat{T}^4\\
&- 2.26782 \times 10^{19} \hat{T}^5
- 1.0404 \times 10^{25} \hat{T}^6
- 9.32122 \times 10^{30} \hat{T}^7+{\cal O}\left(\hat T^{8}\right)
\end{split}
\eeq

\noi
while the $\nu^{10}$ term gives
\beq
\begin{split}
\left.\frac{G_R}{G_{R}^0}\right|_{\nu^{10}}=& \ 84.3425 \hat{T}^2
+ 1.42249 \times 10^6 \hat{T}^3
+ 3.87789 \times 10^{10} \hat{T}^4
+ 3.01096 \times 10^{15} \hat{T}^5\\
&+ 6.16188 \times 10^{20} \hat{T}^6
+ 2.84528 \times 10^{26} \hat{T}^7+{\cal O}\left(\hat T^{8}\right).
\end{split}
\eeq
The analytic formula of the low temperature expansion of the propagator is 

\beq
\label{analyticlowtempprop}
\begin{split}
\frac{G_R}{G_{R}^0}=&1+\sum_{l=1}^{10} (-\pi^4)^l C_l \frac{\Gamma[3+\nu]}{\Gamma[-d_l+\nu]}P_l(\nu)\hat T^{l} +{\cal O}\left(\hat T^{11}\right),
\end{split}
\eeq
where 
\beq
d_l=\frac{4l-1+(-1)^{l+1}}{2}, \quad C_l=2^{e_{l,2}}\prod_{p \text{ odd prime}} p^{-e_{l,p}}
\eeq
with 
\begin{equation}
\begin{split}
    e_{l,2}&=\sum_{k\geq 1}\left(\left\lfloor \frac{2l}{2^k}\right\rfloor-2\left\lfloor \frac{l}{2^k}\right\rfloor\right), e_{l,3}=l+\sum_{k\geq 1} \left\lfloor \frac{l}{3^k}\right\rfloor, e_{l,5}=l+\sum_{k\geq 1} \left\lfloor \frac{l}{5^k}\right\rfloor,\\
    e_{l,p}&=\left\lfloor \frac{l}{\lfloor (p+1)/4\rfloor}\right\rfloor,\quad  p\geq 7.
    \end{split}
\end{equation}
The $P_l(\nu)$ are polynomials with degree $\text{deg} P_l(\nu)=2\lfloor (3l-2)/2\rfloor$ and here we give the first few (more can be found in Appendix \ref{appendixPlnu}),

\beq
\label{poly5}
\begin{split}
P_1(\nu)=&1,\\ P_2(\nu)=&1000 + 324 \nu - 62 \nu^2 - 9 \nu^3 + 7 \nu^4,\\
P_3(\nu)=&4561360 + 
 4318992 \nu + 1416624 \nu^2 + 166500 \nu^3 + 8535 \nu^4 + 
 5148 \nu^5 + 1001 \nu^6,\\ P_4(\nu)=&1701277027200 + 1857907875520\nu + 
 740735545504 \nu^2 + 110670770480 \nu^3\\ &- 3496530760 \nu^4 - 
 1981950060 \nu^5 + 97325502 \nu^6 + 23044145 \nu^7 - 
 5114365 \nu^8\\ & - 85085 \nu^9 + 119119 \nu^{10},\\
 P_5(\nu)=& 3370482566457600 + 4815379612782080 \nu + 
 2888959672781312 \nu^2 \\ & + 933753289342592 \nu^3 + 
 170641909496096 \nu^4 + 16292158804640 \nu^5 \\ & + 
 619660529696 \nu^6 + 44793536536 \nu^7 + 13694280493 \nu^8 + 
 1029635080 \nu^9 \\ &+ 41845942 \nu^{10} + 20692672 \nu^{11} + 
 2263261 \nu^{12}.
 \end{split}
 \raisetag{60pt} 
 \eeq
The asymptotic behavior of the series\footnote{For a similar situation in Seiberg-Witten theory 
see for example \cite{Russo:2012kj}.} was also observed in \cite{Ceplak:2024bja} and the series were approximately resummed using a Borel \cite{Buric:2025anb}
or a mixed Pad\'e-Borel transform \cite{Buric:2025fye}. 
In a similar spirit we will analyse here the low-temperature expansion of the propagator by employing Pad\'e approximants as an alternative resummation technique and compare the result with the numerical
solutions of the Heun equation\footnote{We thank Sa\v so Grozdanov for suggesting it.}. 

The low-temperature expansion of the propagator is an asymptotic series with zero radius of convergence. While truncating this expansion produces a polynomial that gives a reasonable approximation within a narrow domain, it cannot reproduce the correct analytic structure. In contrast, constructing a Pad\'e approximant from the same series yields a rational function that typically provides a more accurate representation over a broader range, since it can capture features such as poles that polynomials miss.

We construct rational approximations of the propagator for $\nu=0.5$ starting from its low-temperature expansion, where we truncated the series \eqref{analyticlowtempprop} at order $d=7$ in $\hat{T}$, yielding a polynomial expansion.
From this truncated series, Padé approximants of order $[n/m]$ with $n+m = d$ of $G_R/G_R^0$ are defined as rational functions of the form
\beq
P_{n,m}(\hat{T}) = \frac{a_0 + a_1 \hat{T} + \cdots + a_n \hat{T}^n}{1 + b_1 \hat{T} + \cdots + b_m \hat{T}^m},
\eeq
whose Taylor expansion around $\hat{T} \to 0$ coincides with the truncated series up to $\mathcal{O}(\hat{T}^{d})$. 

\begin{figure}[htbp]
    \centering
    \begin{subfigure}[b]{0.48\linewidth}
        \centering
        \includegraphics[width=\linewidth]{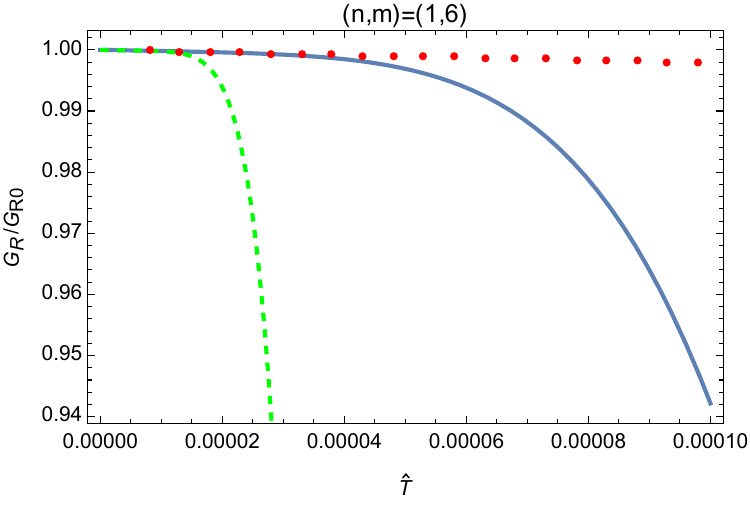}
    \end{subfigure}
    \hfill
    \begin{subfigure}[b]{0.48\linewidth}
        \centering
        \includegraphics[width=\linewidth]{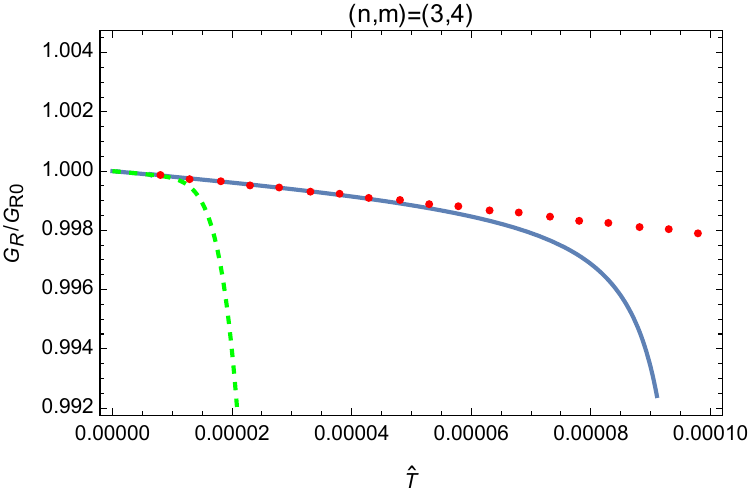}
    \end{subfigure}
    \hfill
    \begin{subfigure}[b]{0.48\linewidth}
        \centering
        \includegraphics[width=\linewidth]{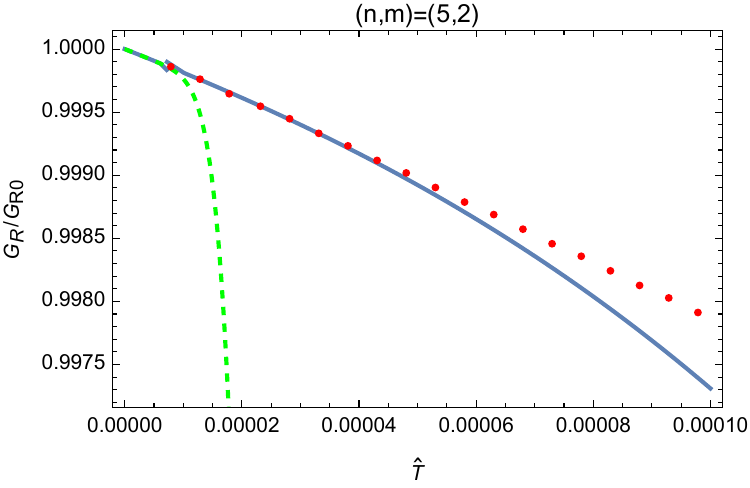}
    \end{subfigure}
    \caption{For $\nu=0.5$, we compare the Pad\'e approximant (blue) with the truncated low-temperature expansion of the propagator (dashed, green) as $T \to 0$, and with the numerical solution (dotted, red) within a finite interval. The three panels display Pad\'e approximants of degree $(n,m) = (1,6)$, $(3,4)$, and $(5,2)$, respectively.
  }
    \label{fig1}
\end{figure}

By construction, the Padé approximants reproduce the asymptotic behavior in the limit $\hat{T} \to 0$. To assess their validity at finite $\hat{T}$, we compared $P_{n,m}(\hat{T})$ against the numerical solution of the propagator. 
We show the matching of the Pad\'e approximant with the truncated perturbative series and the numerical solution in Fig.\ref{fig1}. 
The figure illustrates that the Pad\'e approximant coincides with the truncated perturbative series for $0\leq\hat T < \hat T_1$, while it also agrees with the numerical solution within a finite interval $\hat{T}_1 \leq \hat{T} \leq \hat{T}_2$. Consequently, the Pad\'e approximant provides the appropriate description for $0 < \hat{T} \leq \hat{T}_2$, whereas the truncated perturbative series is valid only for small enough $\hat T<\hat T_1$ (the full 
asymptotic series would be valid only 
strictly at $T=0$) and only the numerical solution must be employed for $\hat{T} > \hat{T}_2$. The three Pad\'e approximants shown are: 
\beq
\begin{split}
&P_{1,6}(\hat{T})=(1 - 8.74285\times 10^{5} \, \hat{T})
(1 - 8.74266 \times 10^5 \, \hat{T} - 1.59192 \times 10^7 \, \hat{T}^2\\ & - 
4.19358 \times 10^{10} \, \hat{T}^3 - 6.35033 \times 10^{14} \, \hat{T}^4 - 
3.79428 \times 10^{19} \, \hat{T}^5 - 4.6629 \times 10^{24} \, \hat{T}^6)^{-1}
\end{split}
\eeq
\beq
\begin{split}
&P_{3,4}(\hat{T})=(1 - 1.27419 \times 10^6 \, \hat{T} + 1.86395 \times 10^{11} \, \hat{T}^2 - 1.7809 \times 10^{15} \, \hat{T}^3)\\ &\times (1 - 1.27417 \times 10^6 \, \hat{T} + 1.86372 \times 10^{11} \, \hat{T}^2 - 1.77756 \times 10^{15} \, \hat{T}^3 - 2.43687 \times 10^{16} \, \hat{T}^4)^{-1}
\end{split}
\eeq
\beq
\begin{split}
 &P_{5,2}(\hat{T})=(1 - 1.20999 \times 10^6 \, \hat{T} + 1.48776 \times 10^{11} \, \hat{T}^2 - 2.65893 \times 10^{12} \, \hat{T}^3\\ & - 6.32177 \times 10^{15} \, \hat{T}^4 - 6.06427 \times 10^{19} \, \hat{T}^5)
(1 - 1.20997 \times 10^6 \, \hat{T} + 1.48754 \times 10^{11} \, \hat{T}^2)^{-1}
\end{split}
 \eeq

From Fig.~\ref{fig1}, it appears that the Pad\'e approximant and the numerical solution begin to deviate around $\hat{T} \approx 10^{-4}$. 
Although extending the convergence interval is beyond the scope of this paper, we verified that higher-order Pad\'e approximants yield a better fit. 
In particular, convergence tends to be better for odd values of $n + m$. 
A rough estimate suggests that achieving a satisfactory fit up to $\hat{T} \approx 1$ would require $n + m \approx 1000$.

\section{Conclusion}

We considered the zero-energy and large-momentum  limit (large compared to the temperature) of a holographic conformal field theory on 
$S^1\times \mathbb{R}^3$ dual to a black brane. This limit translates due to the Matone relation to a large adjoint scalar vev $a$ in the vector 
multiplet. Hence, we performed a large $a$-expansion of the thermal propagator and brought it to a compact form \eqref{GRzGR0}. To obtain 
the low-temperature expansion of the propagator, one has to find $a(k)$ through the inverse Matone relation \eqref{aofx}, plug it into the 
propagator, expand for large $k$ and finally put $k\rightarrow k/T$. However, the final missing ingredient on which the propagator depends 
on, are the large $a$ expansion coefficients $\alpha_{2n}(q,\nu)$ of the instanton prepotential. 

In order to compute these coefficients $\alpha_{2n}$, we used the $S$-duality constraints on the instanton prepotential that led to the quasi-modularity of $\alpha_{2n}$ plus a constant piece and matched it with the Zamolodchikov $q$-recursion. This computational method proved to be efficient, allowing us to obtain $\alpha_{2n}$ up to $2n=58$, with the potential to easily extend to higher orders. 

Finally, we provided the analytic formula of the low-temperature expansion of the propagator up to $\mathcal{O}(T^{44})$ (see \eqref{analyticlowtempprop}) and showed numerically that this expansion becomes asymptotic, confirming the results of \cite{Ceplak:2024bja}.

There are a few issues that deserve to be further investigated:

\bet

\item
The finite $\omega$ terms could be added. There are two technical problems here: as we mentioned before, 
now in the propagator it is not only the vev $a$ to be large, but also $\omega$. It is not clear what type of 
expansion should one do. Second, other mass invariants must be included, so that the expansion is now not  
in powers of the Eisenstein series only, but Jacobi theta functions need to be used making the whole expansion. 

\item
The series obtained seem clearly asymptotic. 
It would be nice to understand the asymptotic series better
doing a full resurgence analysis. The neglected exponentially suppressed terms would here presumably play some role.

\item
The different approaches of writing the propagator in terms of electric or magnetic 
variables ($a$ or $a_D$) should somehow be related to the different 
representation of the conformal blocks, i.e. the $s$ and $t$ channels\footnote{We thank Oleg Lisovvy for several discussions on similar issues.}. 

\eet

\subsubsection*{Acknowledgments}
It is a pleasure to thank Sa\v so Grozdanov and Adrian Lugo for many illuminating discussions on subjects relating this work. We thank them both and Andrei Parnachev for reading the manuscript and comments. The authors acknowledge the financial support from the Slovenian 
Research Agency (research core funding No.~P1-0035). 

\appendix

\section{Holographic two-point functions: black brane}
\label{BHBB}
Let us consider a holographic $4d$ CFT at finite temperature which is dual to AdS$_5$ with a black brane\cite{Son:2002sd}. The black brane metric is 
\beq
ds^2=-f(r)dt^2+f(r)^{-1}dr^2+r^2 d\vec{x}^2,
\eeq
where the redshift factor is \beq
f(r)=r^2\left(1-\left(\frac{R_h}{r}\right)^4\right)
\eeq
with AdS radius $L=1$ and the horizon radius of the black brane $R_h$. The Hawking temperature of the black brane is then 
\beq
T=\frac{R_h}{\pi}.
\eeq  
In order to calculate the two-point function of a scalar operator $\mathcal{O}(x)$ with conformal dimension $\Delta=2+\nu$ in the $4d$ CFT, we can first relate it to its dual scalar field $\phi(x)$ in the bulk with mass $m=\sqrt{\nu^2-4}$ and then solve the wave equation on the black brane background,
\beq
(\square-m^2)\phi=0.
\eeq
Since the scalar wave equation in planar AdS$_5$ commutes with the generators of time translations and the translations along $\vec{x}\in\mathbb{R}^3$, the field can be expanded in a basis of frequency eigenmodes $e^{-i\omega t}$ and momentum eigenmodes $e^{i\vec{k} \vec{x}}$:
\beq
\phi(t,r,\vec{x})=\int d\omega\, d^3k \, e^{-i\omega t+i \vec{k}\cdot \vec{x}} \, \varphi_{\omega,\vec{k}}(r).
\eeq
Then the wave equation reduces to 
\beq
\label{waveeq}
\left( \frac{1}{r^3}\partial_r(r^3 f(r)\partial_r)+\frac{\omega^2}{f(r)}-\frac{k^2}{r^2}-(\nu^2-4)\right)\varphi_{\omega,\vec{k}}(r)=0
\eeq
and we impose ingoing boundary conditions at the horizon, 
\beq
\label{ingoing}
\varphi_{\omega,\vec{k}}(r)\sim (r-R_h)^{-i\omega/f'(R_h)}, \quad r\rightarrow R_h.
\eeq
Near the AdS boundary $r\rightarrow \infty$ this solution behaves as 
\beq
\varphi_{\omega,\vec{k}}(r)\sim \mathcal{A}(\omega,|\vec{k}|) r^{\nu-2}+\mathcal{B}(\omega,|\vec{k}|)r^{-(\nu+2)},
\eeq
where $\mathcal{A}$ is the source for $\mathcal{O}(x)$ in the boundary CFT and $\mathcal{B}$ is the response which is proportional to the expectation value $\langle \mathcal{O}(x)\rangle$ induced by that source. 
Therefore, the retarded Green's function, defined as $G_R(t,\vec{x})=-i \theta(t)\langle [\mathcal{O}(t,\vec{x}),\mathcal{O}(0,0)]\rangle$, in momentum space is the ratio
\beq
G_R(\omega,|\vec{k}|)=\frac{\mathcal{B}(\omega,|\vec{k}|)}{\mathcal{A}(\omega,|\vec{k}|)}.
\eeq
The connection problem is solving for ${\cal A}$ and ${\cal B}$ providing the solution satisfies (\ref{ingoing}).

Rewriting the wave equation \eqref{waveeq} with the new coordinate and field variables,
\beq
z=\frac{r^2}{r^2+R_h^2}, \quad \varphi_{\omega,\vec{k}}(r)=\frac{1}{\sqrt{2}}(1-z)^{1/2}\left(\frac{1}{2}-z\right)^{-1/2}z^{-1/2} \ \chi_{\omega,\vec{k}}(z),
\eeq
gives the Heun equation \eqref{heunwiththetas} with parameters \eqref{BBpar}, where $z=t$ is the horizon and $z=1$ is the boundary. Note that in doing this transformation we absorbed the temperature $T$ in the energy and momentum, i.e.\ $\vec{k}/T\rightarrow \vec{k}$ and $\omega/T\rightarrow \omega$, such that $\omega$ and $\vec{k}$ in the Heun equation are dimensionless variables.

\section{\label{eisenstein}Eisenstein series and (quasi-) modular forms}
The Eisenstein series are defined by
\beq
E_{2k}(\tau)=\frac{1}{2\zeta(2k)}\sum_{(m,n)\in \mathbb{Z}^2\setminus \{(0,0)\}}\frac{1}{(m+n\tau)^{2k}},
\eeq
which absolutely converges to a holomorphic function of $\tau$ in the upper-half plane $\mathbb{H}$.

The Eisenstein series $E_{2k}(\tau)$ for $k>1$ is a modular form of weight $2k$, i.e.\ it is $PSL(2,\mathbb{Z})$-covariant, meaning it transforms as
\beq
\tau \rightarrow \frac{a\tau+b}{c\tau+d}, \quad E_{2k}(\tau)\rightarrow (c\tau+d)^{2k} E_{2k}(\tau),
\eeq
for $a,b,c,d\in\mathbb{Z}$ with $ad-bc=1$.
The Eisenstein series $E_4$ and $E_6$ are the generators of modular forms, meaning that any modular form $M(\tau)$ of weight $2k$ with $k>1$ can be written as
\beq M(\tau)=\sum_{4 n+ 6 m=2k} c_{n,m} E_{4}^{n}(\tau) E_6^m(\tau)\eeq
with $c_{n,m}\in \mathbb{C}$.
The Eisenstein series $E_2$ is not a modular form, it transforms under $PSL(2,\mathbb{Z})$ as
\beq
E_2(\tau)\rightarrow (c\tau+d)^2E_2(\tau)-\frac{6}{\pi}i c(c\tau+d)
\eeq
and is therefore a quasi-modular form of weight $2$. The ring of quasi-modular forms is given by $\mathbb{C}[E_2,E_4,E_6]$ and therefore any quasi-modular form $Q(\tau)$ of weight $2k$ can be written as
\beq
Q(\tau)=\sum_{2l+4 n+ 6 m=2k} c_{l,n,m} E_2^l(\tau)E_{4}^{n}(\tau) E_6^m(\tau)\eeq
with $c_{l,n,m}\in \mathbb{C}$.

The Fourier expansion of the first three Eisenstein series in terms of $q=\exp(i \pi \tau)$ is
\beq
\begin{split}
E_2(\tau) &=1-24 \sum_{n=1}^\infty \sigma_1(n) q^{2n}=1-24 \sum_{n=1}^\infty \frac{n q^{2n}}{1-q^{2n}},\\
E_{4}(\tau ) &= 1 + 240\sum_{n=1}^{\infty} \sigma_3(n) q^{2n}=1 + 240\sum_{n=1}^{\infty} \frac{n^{3} q^{2n}}{1 - q^{2n}}, \\
E_{6}(\tau ) &= 1 - 504\sum_{n=1}^{\infty}\sigma_5(n) q^{2n}=1 - 504\sum_{n=1}^{\infty} \frac{n^{5} q^{2n}}{1 - q^{2n}},
\end{split}
\eeq
where $\sigma_k(n)=\sum_{d|n} d^k$ is the divisor function.

The Eisenstein series $E_4$ and $E_6$ can be rewritten in terms of theta functions
\beq
\begin{split}
E_{4}(\tau ) &= \frac{1}{2} \left(\theta_2(q)^{8} + \theta_3(q)^{8} + \theta_4(q)^{8} \right), \\
E_{6}(\tau ) &= \frac{1}{2} \sqrt{\frac{\left(\theta_2(q)^{8} + \theta_3(q)^{8} + \theta_4(q)^{8} \right)^{3} - 54(\theta_2(q)\theta_3(q)\theta_4(q))^{8}}{2}}, \\
\end{split}
\eeq
which on the other hand can be rewritten in terms of the Dedekind eta function,
\beq
\theta_2(e^{i \tau \pi}) = \frac{2\eta^2(2\tau)}{\eta(\tau)},\quad
\theta_3(e^{i \tau \pi}) = \frac{\eta^5(\tau)}{\eta^2\left(\frac{1}{2}\tau\right)\eta^2(2\tau)}, \quad 
\theta_4(e^{i \tau \pi}) = \frac{\eta^2\left(\frac{1}{2}\tau\right)}{\eta(\tau)}.
\eeq
The Dedekind eta function has the following special values
\beq
\eta(i)=\frac{\Gamma \left(\frac14\right)}{2 \pi ^\frac34}, \quad
\eta\left(\tfrac{1}{2}i\right)=\frac{\Gamma \left(\frac14\right)}{2^\frac78 \pi ^\frac34}, \quad
\eta(2i)=\frac{\Gamma \left(\frac14\right)}{2^\frac{11}{8} \pi ^\frac34}.
\eeq
In the final propagator formula, we have to take $t=1/2$, which means $q=\exp(-\pi)$ and hence $\tau=i$. Therefore, calculating $E_2(i)$ directly and using the special values of the Dedekind eta function for $E_4(i)$ and $E_6(i)$, we obtain 
\beq
E_2(i)=\frac{3}{\pi}, \quad 
E_{4}(i ) = \frac{3}{4}\frac{\pi^2}{\Gamma[3/4]^8}, \quad
E_{6}(i) = 0.
\eeq
In the propagator formula, we also have to use the derivatives of quasi-modular forms. However, this will again give us quasi-modular forms, since the ring of quasi-modular forms is closed under differentiation, especially, we have
\beq
q\frac{d E_2}{d q}=\frac{E_2^2-E_4}{6}, \quad q\frac{d E_4}{d q}=2\frac{E_2 E_4-E_6}{3}, \quad q\frac{d E_6}{d q}=E_2 E_6-E_4^2.
\eeq
Note that although $E_6=0$ for $q=\exp(-\pi)$, we still have to calculate keep it in the calculation of $\tilde{\alpha}_{2n}$, since $q\partial_q E_6\neq0$.

\section{$\tilde{\alpha}_{2n}$ up to $2n=18$}
\label{appendixalphatilde}
Here we give the quasi-modular forms $\tilde{\alpha}_{2n}$ up to $2n=18$ (the first four can be found in \eqref{alphtil8}).
For better readability we write 
\beq
\tilde{\alpha}_{2n}=\frac{(\nu^2-1)^2}{9\cdot 2^{n+2}(2n)!!\  n(n+1)}\times \gamma_{2n},
\eeq
with the following $\gamma_{2n}$:

\begin{equation}
\begin{split}
&\gamma_{10}=\frac{1}{14} \Bigl(
6160 E_2^3 E_4 (\nu^2 - 13) (\nu^2 - 1)^3
+ 5390 E_2^5 (\nu^2 - 1)^4 \\
&+ 165 E_2^2 E_6 (\nu^2 - 1)^2 (11 \nu^4 - 454 \nu^2 + 3323) \\
&+ 88 E_2 E_4^2 (17 \nu^8 - 896 \nu^6 + 13314 \nu^4 - 64328 \nu^2 + 51893) \\
&+ 7 E_4 E_6 (37 \nu^8 - 3556 \nu^6 + 101454 \nu^4 - 1067908 \nu^2 + 3734773)
\Bigr)
\end{split}
\end{equation}

\begin{equation}
\begin{split}
&\gamma_{12}=\frac{1}{12870} \Bigl(
2477475 E_2^4 E_4(\nu^2 - 13)(\nu^2 - 1)^4
+ 1651650 E_2^6 (\nu^2 - 1)^5 \\
&+ 78650 E_2^3 E_6(\nu^2 - 1)^3 (11 \nu^4 - 454 \nu^2 + 3323) \\
&+ 9438 E_2^2 E_4^2 (\nu^2 - 1)^2 (109 \nu^6 - 5463 \nu^4 + 75975 \nu^2 - 312541) \\
&+ 4 E_6^2(3313 \nu^{10} - 529637 \nu^8 + 27763210 \nu^6 - 612870058 \nu^4 
+ 5723061269 \nu^2 - 18823188097) \\
&+ 78  E_2 E_4 E_6 (4127 \nu^{10} - 390883 \nu^8 + 11255510 \nu^6 
- 124152422 \nu^4 + 506385211 \nu^2 - 393101543) \\
&+ 5 E_4^3 (11405 \nu^{10} - 1372285 \nu^8 + 57069362 \nu^6 
- 1064136746 \nu^4 + 8901924337 \nu^2 - 27206840585)
\Bigr)
\end{split}
\raisetag{90pt} 
\end{equation}

\begin{equation}
\begin{split}
&\gamma_{14}=\frac{1}{19305} \Bigl(
20040020 E_2^5 E_4 (\nu^2 - 13)(\nu^2 - 1)^5
+ 10735725 E_2^7 (\nu^2 - 1)^6 \\
&+ 715715 E_2^4 E_6 (\nu^2 - 1)^4 (11 \nu^4 - 454 \nu^2 + 3323) \\
&+ 416416 E_2^3 E_4^2 (\nu^2 - 1)^3 (29 \nu^6 - 1413 \nu^4 + 19335 \nu^2 - 78431) \\
&+ 1183 E_2^2 E_4 E_6 (\nu^2 - 1)^2 (4369 \nu^8 - 400132 \nu^6 + 11084838 \nu^4 - 114436996 \nu^2 + 394051921) \\
&+ 44 E_4^2 E_6 (7907 \nu^{12} - 1436406 \nu^{10} + 94161681 \nu^8 - 2973186148 \nu^6 + 47987533077 \nu^4 \\ &- 371551350726 \nu^2 + 1087272993815) \\
&+ E_2 (\nu^2 - 1) \Bigl(
   9 E_6^2 (43151 \nu^{10} - 6751659 \nu^8 + 350092950 \nu^6 - 7679205206 \nu^4\\ & + 71443709643 \nu^2 - 234419568879) + 28 E_4^3 (62459 \nu^{10} - 7228363 \nu^8 + 293679134 \nu^6 \\ &- 5400861542 \nu^4 + 44812474807 \nu^2 - 136264849055)
\Bigr)
\Bigr)
\end{split}
\raisetag{80pt} 
\end{equation}
\begin{equation}
\begin{split}
&\gamma_{16}=\frac{1}{765765} \Bigl(
4769524760 E_2^6 E_4 (\nu^2 - 13)(\nu^2 - 1)^6
+ 2129252125 E_2^8 (\nu^2 - 1)^7 \\
&+ 185825640 E_2^5 E_6 (\nu^2 - 1)^5 (11 \nu^4 - 454 \nu^2 + 3323) \\
&+ 92912820 E_2^4 E_4^2 (\nu^2 - 1)^4 (41 \nu^6 - 1947 \nu^4 + 26235 \nu^2 - 104969) \\
&+ 1299480 E_2^3 E_4 E_6 (\nu^2 - 1)^3 (1537 \nu^8 - 137836 \nu^6 + 3766974 \nu^4 - 38530108 \nu^2 + 131667433) \\
&+ 3264 E_2 E_4^2 E_6 (\nu^2 - 1)(109856 \nu^{12} - 19321257 \nu^{10} + 1240980357 \nu^8 - 38695339066 \nu^6 \\
&+ 620069070666 \nu^4 - 4782152083053 \nu^2 + 13963525196897) \\
&+ 7644 E_4 E_6^2 (2947 \nu^{14} - 770581 \nu^{12} + 75452223 \nu^{10} - 3724724201 \nu^8 + 100656451049 \nu^6 \\
& - 1478261187903 \nu^4 + 10797325840405 \nu^2 - 30437398423939) \\
&+ 65 E_4^4 (539015 \nu^{14} - 116540465 \nu^{12} + 9815323155 \nu^{10} - 432713630965 \nu^8 \\
& + 10792646612341 \nu^6  - 150127472234643 \nu^4 + 1058779988148401 \nu^2 - 2917721545074695) \\
&+ 68 E_2^2 (\nu^2 - 1)^2 \Bigl(
     98 E_4^3 (146057 \nu^{10} - 16299097 \nu^8 + 647502602 \nu^6 - 11746943330 \nu^4 \\ &+ 96678506077 \nu^2 - 292491603509) + 15 E_6^2 (202831 \nu^{10} - 31105259 \nu^8 + 1596096790 \nu^6\\ & - 34792123606 \nu^4 + 322501955723 \nu^2 - 1055677586479)
\Bigr)
\Bigr)
\end{split}
\raisetag{115pt} 
\end{equation}
\begin{equation}
\begin{split}
&\gamma_{18}=\frac{1}{43648605} \Bigl(
1800650451600 E_2^7 E_4 (\nu^2 - 13)(\nu^2 - 1)^7
+ 687748436375 E_2^9 (\nu^2 - 1)^8 \\
&+ 75027102150 E_2^6 E_6 (\nu^2 - 1)^6 (11 \nu^4 - 454 \nu^2 + 3323) \\
&+ 138511573200 E_2^5 E_4^2 (\nu^2 - 1)^5 (13 \nu^6 - 603 \nu^4 + 8007 \nu^2 - 31609) \\
&+ 224856450 E_2^4 E_4 E_6 (\nu^2 - 1)^4 (4853 \nu^8 - 426884 \nu^6 + 11517006 \nu^4 - 116743652 \nu^2 + 395952677) \\
&+ 1581408 E_2^2 E_4^2 E_6 (\nu^2 - 1)^2 (222386 \nu^{12} - 37938867 \nu^{10} + 2389066857 \nu^8 - 73577606926 \nu^6 \\
&+ 1170615608196 \nu^4 - 8992696408623 \nu^2 + 26200748208977) \\
&+ 91 E_6^3 (9098551 \nu^{16} - 3318610904 \nu^{14} + 470276428132 \nu^{12} - 35035373767592 \nu^{10}\\
& + 1502228813457130 \nu^8 - 37509603645438056 \nu^6 + 525335751790478884 \nu^4\\
& - 3726992758721115032 \nu^2 + 10313477739545468887) \\
&+ 105 E_4^3 E_6 (135775123 \nu^{16} - 40342801304 \nu^{14} + 4774655759380 \nu^{12} - 305532216130856 \nu^{10}\\
& + 11600359659800626 \nu^8 - 264210820184555816 \nu^6 + 3463480616917047316 \nu^4\\
& - 23489837667445792664 \nu^2 + 63057368461013102995) \\
&+ 65892 E_2^3 (\nu^2 - 1)^3 \Bigl(
25 E_6^2 (84593 \nu^{10} - 12731221 \nu^8 + 646710122 \nu^6 - 14011129514 \nu^4\\
& + 129402887221 \nu^2 - 422586845201) + 28 E_4^3 (370639 \nu^{10} - 39975359 \nu^8 + 1554065254 \nu^6 \\
&- 27819098830 \nu^4 + 227110528859 \nu^2 - 683639503363)
\Bigr) \\
&+ 171 E_2 E_4 (\nu^2 - 1) \Bigl(
40 E_4^3 (9656989 \nu^{14} - 2008640383 \nu^{12} + 165036951057 \nu^{10} - 7164021007451 \nu^8\\
& + 177067386656567 \nu^6 - 2450740334488221 \nu^4 + 17238167159967067 \nu^2 - 47436358428035785) \\
&+ 7 E_6^2 (34431581 \nu^{14} - 8758125803 \nu^{12} + 843162713889 \nu^{10} - 41188545488023 \nu^8\\
&+ 1106248892660407 \nu^6- 16191938269934529 \nu^4 + 118057239602068235 \nu^2\\
& - 332484090889525757)
\Bigr)
\Bigr)
\end{split}
\raisetag{125pt} 
\end{equation}

\section{$P_l(\nu)$ up to $l=10$}
\label{appendixPlnu}

We give the polynomials $P_l(\nu)$ from the low-temperature expansion of the propagator in \eqref{analyticlowtempprop}, up to $l=10$ (the first five are already given in \eqref{poly5}).
\beq
\begin{split}
P_6(\nu)=&1449248566558429728000000 + 2202508060056763452748800 \nu \\ & + 
 1422679679687090205463040 \nu^2+ 
 502058430492773141880576 \nu^3\\ & + 
 101304014393631422420992 \nu^4 + 10250363050169528269440\nu^5\\ & + 
 43769734212643252800 \nu^6 - 90640025345532672288 \nu^7\\ & - 
 3799708109870825656\nu^8 + 632172469553364420\nu^9 + 
 24422937301909930 \nu^{10}\\ & - 4025662418725677 \nu^{11}+ 
 155598706432911 \nu^{12} + 17835774403590 \nu^{13}\\ & - 
 5959405472020 \nu^{14} - 66994788861 \nu^{15} + 52107058003 \nu^{16}
\end{split}
\raisetag{90pt} 
\eeq

\beq
\begin{split}
P_7(\nu)=&1298649823891694771673600000 + 2274588609510698838781440000 \nu\\ & + 
 1765870706501114273978470400 \nu^2 + 
 799386677935937946478648320 \nu^3 \\ &+ 
 232650575609538014385083648 \nu^4 + 
 44976822119182232773974528 \nu^5 \\ &+ 
 5694023931629785610664704 \nu^6 + 
 435027736240286972784000 \nu^7\\ & + 15510110215415671889456 \nu^8 + 
 200119127975079578736 \nu^9\\ & + 81900351486218166608 \nu^{10} + 
 11492453942992100940 \nu^{11} \\ &+ 442814004978174719\nu^{12} + 
 10126921617057444 \nu^{13} + 2610383651583347 \nu^{14} \\ &+ 
 88059701326740 \nu^{15} - 216303410323 \nu^{16} + 
 2590465169292 \nu^{17} + 215872097441\nu^{18}
\end{split}
\raisetag{90pt} 
\eeq

\beq
\begin{split}
&P_8(\nu)=308458337180287001104607725670400000 \\ &+ 
 560837637780705338243461708498944000 \nu + 
 454559273058931015393692653656268800 \nu^2 \\ &+ 
 216187294110503718808290692414115840 \nu^3+ 
 66552579148342992077152707060881408 \nu^4 \\ &+ 
 13690863858192997167275810885710848 \nu^5 + 
 1841749628537456412420845254915584 \nu^6\\ & + 
 142183869814189486712805662557440 \nu^7 + 
 2230691307854165199403125237888 \nu^8\\ & - 
 641766544158252583424026251072 \nu^9 - 
 43851191463294404538206753376 \nu^{10}\\ & + 
 1597067263815762818232494640 \nu^{11} + 
 227427303719187295114847688 \nu^{12}\\ & - 
 5209229097118050071989572 \nu^{13} - 
 900630229074100795863726\nu^{14} \\ &+ 
 31404743366506609679415 \nu^{15} + 2489137755514699057953 \nu^{16}\\ & - 
 131741782723570172157 \nu^{17} + 10873466630151660369 \nu^{18} + 
 430657623921690165 \nu^{19} \\ &- 189830149584512437 \nu^{20} - 
 2389056502379547 \nu^{21} + 796352167459849 \nu^{22}
\end{split}
\raisetag{90pt} 
\eeq

\beq
\begin{split}
&P_9(\nu)=601631416585763948100869522036739072000000 \\ &+ 
 1200307677168967367629861100982154199040000 \nu\\ & + 
 1089207702739811084207822755241458499584000 \nu^2\\ & + 
 595681976230185077872756542689821365043200 \nu^3\\ & + 
 218973663443352144179924430356065024983040 \nu^4\\ & + 
 57009114934608848909950136646063506669568 \nu^5\\ & + 
 10754735305539390962180572178535251623936 \nu^6\\ & + 
 1470705081729382370763534205165493809152 \nu^7\\ & + 
 142008862152667112331829277669683137024 \nu^8\\ & + 
 8976763775386750149828712979122019328 \nu^9 \\ &+ 
 297707018061182614478218366667667456 \nu^{10} + 
 1414373118639550135124682805339392 \nu^{11}\\ & + 
 297463685619804203814165927501504 \nu^{12} + 
 77863747716988002869604128284608 \nu^{13}\\ & + 
 4422825265001119238391924110016 \nu^{14} + 
 39102932060132643841761096912 \nu^{15}\\ & + 
 2663703298976670574393134819 \nu^{16} + 
 797519474403812849515764528 \nu^{17}\\ &+ 
 32928608806916832880414356 \nu^{18} + 
 464363903334125957125392 \nu^{19} \\ &+ 
 69879799119612388317554 \nu^{20} - 2649125946635594458032\nu^{21} \\ &- 
 324098494996046534764 \nu^{22} + 57785133504412265952 \nu^{23} + 
 4213499318030061059 \nu^{24}
\end{split}
\raisetag{160pt} 
\eeq

\beq
\begin{split}
&P_{10}(\nu)=543311270349889057424461200870008883717120000000000\\ & + 
 1112553120273948652717033815390331385777275699200000 \nu \\ &+ 
 1039709889350257667676301619392150589037128663040000 \nu^2\\ & + 
 587720631376690468398528015010800713337425059840000 \nu^3 \\ &+ 
 224183392519402763842077949359384187583550103552000 \nu^4\\ & + 
 60810772617745044421683688608513323605738298081280 \nu^5\\ & + 
 11997533100179304955183788849171136657427677741056 \nu^6\\ & + 
 1718756442219155234402831309323683177452485656576 \nu^7 \\ &+ 
 172623438551462056556012010423396011656180789248 \nu^8\\ & + 
 10782885816203721418523648572529952102172620800 \nu^9\\ & + 
 208144664048085999479900194465772465500421120 \nu^{10}\\ & - 
 27033883781491793153801040748069806559342080 \nu^{11}\\ & - 
 2166314968820198328049738378644260063984640 \nu^{12}\\ & + 
 2202298964749622030867625963347316000000 \nu^{13}\\ &+ 
 6804603457698135755133319271482871544960 \nu^{14}\\ & + 
 126721791237026657178873295998176794560 \nu^{15}\\ & - 
 17822331928400743384179310227493216920 \nu^{16}\\ & - 
 396737797611816495544078224136475100 \nu^{17}\\ & + 
 49234710018788367898621636123209090 \nu^{18} + 
 535933227945055560389326589687415 \nu^{19}\\ & - 
 135451191733609956971407350029505 \nu^{20} + 
 731680383840174001070596273100 \nu^{21}\\ & + 
 213634378328808167245098519680 \nu^{22} - 
 3850546266076465434078098070 \nu^{23} \\ &+ 
 1231918522152644589733560090 \nu^{24} + 
 26520318343218322627924920 \nu^{25}\\ & - 
 9793650439946554376100906 \nu^{26}- 
 134574954718562120163401 \nu^{27}\\ & + 22976211781217922954727 \nu^{28}
\end{split}
\raisetag{180pt} 
\eeq


\begin{thebibliography}{99}

\bibitem{Maldacena:1997re}
J.~M.~Maldacena,
``The Large $N$ Limit of Superconformal Field Theories and Supergravity,''
Adv. Theor. Math. Phys. \textbf{2} (1998), 231-252
[arXiv:hep-th/9711200 [hep-th]].

\bibitem{Gubser:1998bc}
S.~S.~Gubser, I.~R.~Klebanov and A.~M.~Polyakov,
``Gauge Theory Correlators from Noncritical String Theory,''
Phys. Lett. B \textbf{428} (1998), 105-114
[arXiv:hep-th/9802109 [hep-th]].

\bibitem{Witten:1998qj}
E.~Witten,
``Anti-de~Sitter Space and Holography,''
Adv. Theor. Math. Phys. \textbf{2} (1998), 253-291
[arXiv:hep-th/9802150 [hep-th]].

\bibitem{Grozdanov:2019uhi}
S.~Grozdanov, P.~K.~Kovtun, A.~O.~Starinets and P.~Tadi\'c,
``The Complex Life of Hydrodynamic Modes,''
JHEP \textbf{11} (2019), 097
[arXiv:1904.12862 [hep-th]].

\bibitem{Policastro:2001yb}
G.~Policastro and A.~Starinets,
``On the Absorption by Near Extremal Black Branes,''
Nucl. Phys. B \textbf{610} (2001), 117-143
[arXiv:hep-th/0104065 [hep-th]].

\bibitem{Rodriguez-Gomez:2021pfh}
D.~Rodriguez-G\'omez and J.~G.~Russo,
``Correlation Functions in Finite Temperature CFT and Black Hole Singularities,''
JHEP \textbf{06} (2021), 048
[arXiv:2102.11891 [hep-th]].

\bibitem{Rodriguez-Gomez:2021mkk}
D.~Rodriguez-G\'omez and J.~G.~Russo,
``Thermal Correlation Functions in CFT and Factorization,''
JHEP \textbf{11} (2021), 049
[arXiv:2105.13909 [hep-th]].

\bibitem{Fitzpatrick:2019zqz}
A.~L.~Fitzpatrick and K.~W.~Huang,
``Universal Lowest-Twist in CFTs from Holography,''
JHEP \textbf{08} (2019), 138
[arXiv:1903.05306 [hep-th]].

\bibitem{Parisini:2022wkb}
E.~Parisini, K.~Skenderis and B.~Withers,
``Embedding Formalism for CFTs in General States on Curved Backgrounds,''
Phys. Rev. D \textbf{107} (2023) no.6, 066022
[arXiv:2209.09250 [hep-th]].


\bibitem{Bajc:2022wws}
B.~Bajc and A.~R.~Lugo,
``Holographic Thermal Propagator for Arbitrary Scale Dimensions,''
JHEP \textbf{05} (2023), 103
[arXiv:2212.13639 [hep-th]].

\bibitem{Huang:2024wbq}
K.~W.~Huang,
``Resummation of Multistress Tensors in Higher Dimensions,''
Phys. Rev. D \textbf{111} (2025) no.4, 046016
[arXiv:2406.07458 [hep-th]].

\bibitem{Buric:2025fye}
I.~Buri{\'c}, I.~Gusev and A.~Parnachev,
``Holographic Correlators from Thermal Bootstrap,''
[arXiv:2508.08373 [hep-th]].

\bibitem{El-Showk:2011yvt}
S.~El-Showk and K.~Papadodimas,
``Emergent Spacetime and Holographic CFTs,''
JHEP \textbf{10} (2012), 106
[arXiv:1101.4163 [hep-th]].

\bibitem{Iliesiu:2018fao}
L.~Iliesiu, M.~Kolo{\u{g}}lu, R.~Mahajan, E.~Perlmutter and D.~Simmons-Duffin,
``The Conformal Bootstrap at Finite Temperature,''
JHEP \textbf{10} (2018), 070
[arXiv:1802.10266 [hep-th]].

\bibitem{Buric:2025anb}
I.~Buri{\'c}, I.~Gusev and A.~Parnachev,
``Thermal Holographic Correlators and Kms Condition,''
[arXiv:2505.10277 [hep-th]].

\bibitem{Niarchos:2025cdg}
V.~Niarchos, C.~Papageorgakis, A.~Stratoudakis and M.~Woolley,
``Deep Finite Temperature Bootstrap,''
[arXiv:2508.08560 [hep-th]].

\bibitem{Bonelli:2022ten}
G.~Bonelli, C.~Iossa, D.~Panea Lichtig and A.~Tanzini,
``Irregular Liouville Correlators and Connection Formulae for Heun Functions,''
Commun. Math. Phys. \textbf{397} (2023) no.2, 635-727
[arXiv:2201.04491 [hep-th]].

\bibitem{Dodelson:2022yvn}
M.~Dodelson, A.~Grassi, C.~Iossa, D.~Panea Lichtig and A.~Zhiboedov,
``Holographic Thermal Correlators from Supersymmetric Instantons,''
SciPost Phys. \textbf{14} (2023) no.5, 116
[arXiv:2206.07720 [hep-th]].


\bibitem{Nekrasov:2009rc}
N.~A.~Nekrasov and S.~L.~Shatashvili,
``Quantization of Integrable Systems and Four Dimensional Gauge Theories,''
[arXiv:0908.4052 [hep-th]].

\bibitem{Seiberg:1994aj}
N.~Seiberg and E.~Witten,
``Monopoles, Duality and Chiral Symmetry Breaking in ${\mathcal{N}}\!=2$ Supersymmetric QCD,''
Nucl. Phys. B \textbf{431} (1994), 484-550
[arXiv:hep-th/9408099 [hep-th]].

\bibitem{Matone:1995rx}
M.~Matone,
``Instantons and Recursion Relations in ${\mathcal{N}}\!=2$ SUSY Gauge Theory,''
Phys. Lett. B \textbf{357} (1995), 342-348
[arXiv:hep-th/9506102 [hep-th]].

\bibitem{Nekrasov:2002qd}
N.~A.~Nekrasov,
``Seiberg-Witten Prepotential from Instanton Counting,''
Adv. Theor. Math. Phys. \textbf{7} (2003) no.5, 831-864
[arXiv:hep-th/0206161 [hep-th]].

\bibitem{Alday:2009aq}
L.~F.~Alday, D.~Gaiotto and Y.~Tachikawa,
``Liouville Correlation Functions from Four-Dimensional Gauge Theories,''
Lett. Math. Phys. \textbf{91} (2010), 167-197
[arXiv:0906.3219 [hep-th]].


\bibitem{Zamolodchikov:1984eqp}
A.~B.~Zamolodchikov,
``Conformal Symmetry in Two-Dimensions: an Explicit Recurrence Formula for the Conformal Partial Wave Amplitude,''
Commun. Math. Phys. \textbf{96} (1984), 419-422.

\bibitem{Zamolodchikov:1987avt}
A.~B.~Zamolodchikov,
``Conformal Symmetry in Two-Dimensional Space: Recursion Representation of Conformal Block,''
Theor. Math. Phys. \textbf{73} (1987) no.1, 1088-1093.

\bibitem{Zamolodchikov:1995aa}
A.~B.~Zamolodchikov and A.~B.~Zamolodchikov,
``Structure Constants and Conformal Bootstrap in Liouville Field Theory,''
Nucl. Phys. B \textbf{477} (1996), 577-605.
[arXiv:hep-th/9506136 [hep-th]].

\bibitem{Poghossian:2009mk}
R.~Poghossian,
``Recursion Relations in CFT and ${\mathcal{N}}\!=2$ Sym Theory,''
JHEP \textbf{12} (2009), 038
[arXiv:0909.3412 [hep-th]].



\bibitem{neven}
Neven Elezovi\'c, 
``Asymptotic expansions of Gamma and related functions, binomial coefficients, inequalities and means,'' 
Journal of Mathematical Inequalities, Volume 9, Number 4 (2015), 1001-1054.

\bibitem{Gamma}
A.~Erd\' elyi and F.~G.~Tricomi, 
``The asymptotic expansion of a ratio of gamma functions,"
Pacific J. Math. 1(1): 133-142 (1951).

\bibitem{Billo:2013jba}
M.~Bill\'o, M.~Frau, L.~Gallot, A.~Lerda and I.~Pesando,
``Modular Anomaly Equation, Heat Kernel and S-Duality in ${\mathcal{N}}\!=2$ Theories,''
JHEP \textbf{11} (2013), 123
[arXiv:1307.6648 [hep-th]].


\bibitem{Billo:2015pjb}
M.~Bill{\'o}, M.~Frau, F.~Fucito, A.~Lerda and J.~F.~Morales,
``S-Duality and the Prepotential in $ \mathcal{N}={2}^{\star } $ Theories (I): the ADE Algebras,''
JHEP \textbf{11} (2015), 024
[arXiv:1507.07709 [hep-th]].

\bibitem{Minahan:1997if}
J.~A.~Minahan, D.~Nemeschansky and N.~P.~Warner,
``Instanton expansions for mass deformed N=4 superYang-Mills theories,''
Nucl. Phys. B \textbf{528} (1998), 109-132
[arXiv:hep-th/9710146 [hep-th]].


\bibitem{Billo:2013fi}
M.~Bill\'o, M.~Frau, L.~Gallot, A.~Lerda and I.~Pesando,
``Deformed ${\mathcal{N}}\!=2$ Theories, Generalized Recursion Relations and S-Duality,''
JHEP \textbf{04} (2013), 039
[arXiv:1302.0686 [hep-th]].

\bibitem{zagier}
D. Zagier: Elliptic modular forms and their applications, first part in ”The 1-2-3 of modular forms”, Universitext. Springer-Verlag, Berlin, 2008. (available at \url{http://people.mpim-bonn.mpg.de/zagier/files/doi/10.1007/978-3-540-74119-0_1/fulltext.pdf})

\bibitem{Russo:2012kj}
J.~G.~Russo,
``A Note on Perturbation Series in Supersymmetric Gauge Theories,''
JHEP \textbf{06} (2012), 038
[arXiv:1203.5061 [hep-th]].

\bibitem{Ceplak:2024bja}
N.~{\v{C}}eplak, H.~Liu, A.~Parnachev and S.~Valach,
``Black Hole Singularity from Ope,''
JHEP \textbf{10} (2024), 105
[arXiv:2404.17286 [hep-th]].

\bibitem{Son:2002sd}
D.~T.~Son and A.~O.~Starinets,
``Minkowski space correlators in AdS / CFT correspondence: Recipe and applications,''
JHEP \textbf{09} (2002), 042
[arXiv:hep-th/0205051 [hep-th]].

\end{thebibliography}

\end{document}